\documentclass[twocolumn,astrosym,apj,]{aastex631}
\usepackage{graphicx,amssymb,amsmath}    
\usepackage{amsthm,amsfonts,amscd,wasysym,cleveref}

\shorttitle{J0749+2255: kpc-separation dual quasar at cosmic noon with JWST}
\shortauthors{Ishikawa et al.}

\newcommand{\progNum}{02654}

\newcommand{\target}{J0749+2255}
\newcommand{\targetfull}{SDSSJ074229.96+225511.7}

\newcommand{\targSW}{J0749+2255-SW}
\newcommand{\targNE}{J0749+2255-NE}

\newcommand{\hst}{\textit{HST}}
\newcommand{\gaia}{\textit{Gaia}}
\newcommand{\jwst}{\textit{JWST}}

\newcommand{\sdss}{\textrm{SDSS}}

\newcommand{\qdfit}{\texttt{q3dfit}}

\newcommand{\python}{\texttt{Python}}
\newcommand{\jwstpipurl}{\url{https://github.com/spacetelescope/jwst}}
\newcommand{\jwstotherpipe}{\url{https://jwst-pipeline.readthedocs.io/en/latest/jwst/cube_build/main.html}}
\newcommand{\qsofiturl}{\url{https://github.com/legolason/PyQSOFit}}
\newcommand{\qdfiturl}{\url{https://q3dfit.readthedocs.io/en/latest/index.html}}

\newcommand\pyqsofit{\texttt{PyQSOFit}}

\newcommand{\oi}{[\textrm{O}~\textsc{i}]}

\newcommand{\oiii}{[\textrm{O}~\textsc{iii}]}
\newcommand{\hei}{\textrm{He}~\textsc{i}}

\newcommand{\nii}{[\textrm{N}~\textsc{ii}]}
\newcommand{\sii}{[\textrm{S}~\textsc{ii}]}

\newcommand{\feii}{\textrm{Fe}~\textsc{ii}}
\newcommand{\mgii}{\textrm{Mg}~\textsc{ii}}

\newcommand{\ha}{${\rm H\alpha}$}
\newcommand{\hb}{${\rm H\beta}$}
\newcommand{\hg}{${\rm H\gamma}$}

\newcommand{\pah}{PAH 3.3 $\mu$m}

\newcommand{\niiha}{[\textrm{N}~\textsc{ii}]/{$\rm H\alpha$}}
\newcommand{\siiha}{[\textrm{S}~\textsc{ii}]/{$\rm H\alpha$}}
\newcommand{\oiha}{[\textrm{O}~\textsc{i}]/{$\rm H\alpha$}}
\newcommand{\oiiihb}{[\textrm{O}~\textsc{iii}]/{$\rm H\beta$}}

\newcommand{\mum}{$\mu$m}
\newcommand{\kms}{km s$^{-1}$}
\newcommand{\ergs}{erg s$^{-1}$}
\newcommand{\ergflux}{erg s$^{-1}$ cm$^{-2}$ \AA$^{-1}$}

\newcommand{\Mdyn}{$M_{dyn}$}
\newcommand{\eddrat}{$\lambda_{Edd}$}
\newcommand{\Ledd}{$L_{Edd}$}
\newcommand{\Mbh}{$M_{\textrm{BH}}$}
\newcommand{\Lbol}{$L_{bol}$}

\newcommand{\HaHb}{${\rm H\alpha}/{\rm H\beta}$}
\newcommand{\fQratio}{$f_{\rm SW}/f_{\rm NE}$}

\begin{document}

\title{VODKA-JWST: Synchronized growth of two SMBHs in a massive gas disk? A 3.8 kpc separation dual quasar at cosmic noon with NIRSpec IFU}

\newcommand{\jhu}{\rm Department of Physics and Astronomy, Johns Hopkins University, Baltimore, MD 21218, USA}
\newcommand{\stsci}{\rm Space Telescope Science Institute, 3700 San Martin Drive, Baltimore, MD 21218, USA}
\newcommand{\ias}{\rm Institute for Advanced Study, Princeton University, Princeton, NJ 08544, USA}
\newcommand{\uiuc}{\rm Department of Astronomy, University of Illinois at Urbana-Champaign, Urbana, IL 61801, USA}
\newcommand{\heidelberg}{\rm Zentrum für Astronomie der Universität Heidelberg, Astronomisches Rechen-Institut, Mönchhofstr 12-14, D-69120 Heidelberg, Germany}
\newcommand{\mki}{\rm MIT Kavli Institute for Astrophysics and Space Research, Massachusetts Institute of Technology, Cambridge, MA 02139, USA}


\correspondingauthor{Yuzo Ishikawa}
\email{yishika2@mit.edu}

\author[0000-0001-7572-5231]{Yuzo Ishikawa}
\affiliation{\jhu}
\affiliation{\mki}

\author[0000-0001-6100-6869]{Nadia L. Zakamska}
\affiliation{\jhu}

\author[0000-0003-1659-7035]{Yue Shen}
\affiliation{\uiuc}

\author[0000-0003-0049-5210]{Xin Liu}
\affiliation{\uiuc}
\affiliation{\rm National Center for Supercomputing Applications, University of Illinois at Urbana-Champaign, Urbana, IL 61801, USA}
\affiliation{\rm Center for Artificial Intelligence Innovation, University of Illinois at Urbana-Champaign, 1205 West Clark Street, Urbana, IL 61801, USA}

\author[0000-0002-9932-1298]{Yu-Ching Chen}
\affiliation{\jhu}
\affiliation{\uiuc}

\author[0000-0003-4250-4437]{Hsiang-Chih Hwang}
\affiliation{\rm School of Natural Sciences, Institute for Advanced Study, 1 Einstein Drive, Princeton, NJ 08540, USA}

\author[0000-0002-0710-3729]{Andrey Vayner}
\affiliation{\jhu}
\affiliation{\rm IPAC, California Institute of Technology, 1200 E. California Boulevard, Pasadena, CA 91125, USA}

\author[0000-0002-1608-7564]{David S. N. Rupke}
\affiliation{\rm Department of Physics, Rhodes College, 2000 N. Parkway, Memphis, TN 38112, USA}
\affiliation{\heidelberg}

\author[0000-0002-3158-6820]{Sylvain Veilleux}
\affiliation{\rm Department of Astronomy and Joint Space-Science Institute, University of Maryland, College Park, MD 20742, USA}

\author[0000-0003-2212-6045]{Dominika Wylezalek}
\affiliation{\heidelberg}

\author[0000-0001-7681-9213]{Arran C. Gross}
\affiliation{\uiuc}

\author[0000-0002-4419-8325]{Swetha Sankar}
\affiliation{\jhu}

\author{Nadiia Diachenko}
\affiliation{\jhu}


\begin{abstract}
The search for dual supermassive black holes (SMBHs) is of immense interest in modern astrophysics. Galaxy mergers may fuel and produce SMBH pairs. Actively accreting SMBH pairs are observed as a dual quasar, which are vital probes of SMBH growth. Dual quasars at cosmic noon are not well characterized. \textit{Gaia} observations have enabled a novel technique to identify dual quasars at kpc scales, based on the small jitters of the light centroid as the two quasars vary stochastically. We present the first detailed study of a $z=2.17$, $0.46\arcsec$, 3.8 kpc separation dual quasar, \target, using JWST/NIRSpec integral field unit spectroscopy. Identified by \textit{Gaia}, \target\ is one of the most distant, small separation dual quasars known. We detect the faint ionized gas of the host galaxy, traced by the narrow \ha\ emission. Line ratios indicate ionization from the two quasars and from intense star formation. Spectral analysis of the two quasars suggests that they have similar black hole properties, hinting at the possible synchronized accretion activity or lensed quasar images. Surprisingly, the ionized gas kinematics suggest a rotating disk rather than a disturbed system expected in a major gas-rich galaxy merger. Numerical simulations show that this is a plausible outcome of a major gas-rich galaxy merger several tens of Myr before coalescence. Whether \target\ reflects an interesting phase of dual quasar evolution or is a lensed quasar remains unclear. Thus, this study underscores the challenges in definitively distinguishing between a dual and lensed quasars, with observations supporting either scenario.
\end{abstract}




\section{Introduction} \label{sec:intro}
Identifying and characterizing dual and binary supermassive black holes (SMBHs) represents a critical frontier in understanding the formation and evolution of galaxies and their central black holes. Galaxy mergers are often invoked as a key mechanism to fuel and produce SMBH pairs \citep{Begelman1980}. Following a merger of two galaxies, the two SMBHs are thought to inspiral into a bound binary through dynamical friction and interaction with the surrounding gas and stars  \citep{Begelman1980, Gould2000, Milosavljevic2001, Blaes2002, Yu2002a}. To investigate this process, it is essential to identify SMBH pairs at various evolutionary stages, corresponding to different separations: from tens of kpcs at the merger onset to $\sim$kpc (dual SMBH) to $\lesssim 10$ pc, where the SMBHs become gravitationally bound as a binary \citep{Colpi2011, Dotti2009, Volonteri2016, DeRosa2019, Pfeifle2024}. Understanding the rate at which SMBHs spiral into the center of the final galactic merger products, as well as the subsequent evolution on pc-scale separations, is important for prospects of low-frequency gravitational waves \citep[e.g.][]{Baker2006, LIGO2016, LISA2017, LISA2024, NANOGrav2018, EPTA2023}.

The galactic inspiraling phase is likely associated with episodes of intense accretion and winds, which act as key agents of galaxy evolution. Such processes, collectively known as `feedback', can heat up or expel gas that would otherwise lead to star formation \citep{Hopkins2006}. Hydrodynamical simulations predict that tidal torques drive substantial gas inflows to the central regions, supplying fuel for accretion onto the SMBHs. When these SMBHs are actively accreting during inspiral, they may be observed as dual/binary quasars or active galactic nuclei (AGNs). There is some evidence that major mergers may play a role in triggering the most luminous quasars, either by directly feeding gas \citep{Barnes1992, Sanders1996, DiMatteo2005, Veilleux2009a, Veilleux2009b} or by triggering an initial growth phase \citep{McAlpine2018} until secular processes dominate SMBH growth \citep[e.g.][]{Kormendy2004, Hopkins2009, Greene2010, Cisternas2011}. Although the exact mechanisms of fueling remain debated \citep[e.g.][]{Mechtley2016}, it is evident that black hole growth requires an abundant supply of cold, dense gas that is somehow transported to the circumnuclear regions \citep{Heckman2014}. 

Identifying and placing unbiased statistical constraints on the population of quasar pairs has not been possible until very recently. Past methods of candidate selection relied on mining photometric quasars catalogs - typically in the optical or X-rays - for spectroscopic follow-up \citep[e.g.][]{Comerford2015, LiuX2018,Stemo2021}, which limited discovery to pairs with large angular separations ($>1\arcsec$ corresponding to $>10\textrm{ kpc}$). Double-peaked emission lines (e.g. \oiii) can originate in dual quasars \citep{LiuX2018}, but they can also be a by-product of quasar-driven galactic winds instead \citep{Shen2011, Fu2012}. Dual AGN/quasars with small separations have been discovered and studied, albeit in small numbers, particularly at $z \lesssim 1$ \citep[e.g.][]{Rodriguez2006, Bianchi2008, Koss2011,Kharb2017, Koss2018, Goulding2019, Silverman2020, Tubin2021, ChenYC2022, Ciurlo2023, Scialpi2024}, where they are easier to detect, observe, and resolve. Yet, little is known about dual quasars at $z\sim2-3$ \citep[e.g.][]{Glikman2023a, Shen2021, Tang2021, Perna2023}, the epoch of peak quasar activity and galaxy formation, due to significant observational challenges. Many dual quasar candidates are often serendipitously discovered. Consequently, the resulting inhomogeneous samples make it difficult to draw concrete comparisons between properties. 

The advent of \gaia\ \citep{Gaia2016} allows for a systematic search for close quasar pairs \citep[e.g.][]{Hwang2020,Shen2019b,Mannucci2022}, even if unresolved by standard imaging methods. The approach of interest in this paper is Varstrometry for Off-nucleus and Dual sub-Kpc AGN (VODKA; \citealt{Shen2019b,Hwang2020}). Since quasars vary stochastically in the rest-frame UV and optical range \citep{Sesar2007}, asynchronous variation in flux in the unresolved components of the pair introduces an astrometric shift in the system's photocenter observed by \gaia\ \citep{LiuY2015, LiuY2016}. The varstrometry method takes these observed \gaia\ astrometric jitters and has successfully identified multiple point-like sources. Follow-up observations have confirmed some of these candidates as physically associated dual quasars at both low and high redshifts \citep{Shen2019b, Hwang2020, Shen2021, ChenYC2022, ChenYC2023a}. Throughout this paper, we refer to the ``varstrometry'' method as VODKA, and collectively refer to dual AGNs and quasars as ``dual quasars''.

\targetfull\ (\target\ henceforth) is one of such VODKA-selected targets. Its dual quasar nature was inferred through follow-up studies conducted with the Hubble Space Telescope (HST), Gemini, Chandra, and VLA \citep{Shen2021, ChenYC2023a}. \target\ was observed with the JWST \citep{Gardner2006} Near-Infrared Spectrograph (NIRSpec; \citealt{Jakobsen2022}) instrument in the integral field unit (IFU; \citealt{Boker2022}) mode on UTC 2022 November 23 as part of the GO Cycle 1 program (ID: \progNum; PI: Ishikawa) to uncover the faint host galaxy and characterize the gas kinematics. In this paper, we present the first spatially resolved spectroscopic observations of a $z=2.17$ sub-arcsec, kpc-scale separation dual quasar with JWST. 

This is one of the first JWST studies of a close-separation dual quasar in the early universe. In Section \ref{sec:prevOBS}, we summarize the known properties of \target. Section \ref{sec:dataRedux} outlines the JWST observations and data reduction. In Section \ref{sec:specAnaly}, we present the spectral analyses of the dual quasars and the extended emission, including the quasar point-spread-function (PSF) subtraction. We end with an extended discussion of the interpretation of the JWST observations in Sections \ref{sec:disc:duallens}, \ref{sec:disc:dualBH}, and \ref{sec:disc:vodka}. In Section \ref{sec:disc:duallens} we compare the evidence for the dual quasar and lensing hypotheses. Then assuming the dual quasar scenario, we discuss the interpretations of the JWST data in Sections \ref{sec:disc:merger}, \ref{sec:disc:syncBH}, including a comparison with low redshift analogs in Section \ref{sec:disc:lowz}. We explore the implications of the VODKA method in Section \ref{sec:disc:vodka}, and finally conclude in Section \ref{sec:concl}. All spectral fits are performed with wavelengths in the vacuum scale. We adopt the $\Lambda$CDM cosmology with $h = 0.7$, $\Omega_M = 0.3$, and $\Omega_{\Lambda} = 0.7$.

\section{Summary of J0749+2255}\label{sec:prevOBS}
\target\ is an optically selected broad-line quasar at $z=2.17$ that was spectroscopically confirmed by the Sloan Digital Sky Survey (\sdss; \citealt{Schneider2010}). Initially identified as a single quasar in SDSS, \target\ was later flagged as a dual quasar candidate with VODKA \citep{Shen2021}. An HST Snapshot program with \hst/F475W+F814W imaging revealed two point-like cores separated by $\sim0.5\arcsec$ corresponding to a physical separation of $\sim3.8\textrm{ kpc}$ \citep{Shen2021, ChenYC2022}.

\cite{ChenYC2023a} conducted extensive multi-wavelength imaging and spectroscopic follow-up observations of \target. Using X-ray observations ($2-8\textrm{ keV}$) with Chandra/ACIS-S and radio imaging (6 and 15 GHz) with the VLA A-config, \cite{ChenYC2023a} revealed two quasar nuclei that are spatially coincident with the HST cores - the optically brighter quasar to the southwest (\targSW) and the fainter quasar to the northeast (\targNE). Keck adaptive-optics-assisted IR/$K_p$-band and \hst/F160W imaging further suggested the indirect detection of an extended host galaxy. PSF modeling of the \hst/F160W imaging revealed extended tidal tail features, indicative of possible ongoing galaxy merger activity. Finally, spatially resolved \hst/STIS spectroscopy and Gemini/GMOS+GNIRS spectroscopy revealed two rest-frame UV and optical quasar spectra, which showed some differences in their continua and emission lines. There was no sign of a foreground galaxy that could have served as a lens to produce the two observed quasars' images. Differences in the observed quasar spectra and the nondetection of a lens galaxy argued against a gravitationally lensed quasar. 

To further test the dual quasar hypothesis and to study the gas dynamics of the host galaxies, we proposed JWST IFU observations (ID: \progNum; PI: Ishikawa) of \target\ in the near-infrared (NIRSpec) and mid-infrared (Mid-Infrared Instrument; MIRI). The companion study \citep{ChenYC2024} reports observations with MIRI Medium Resolution Spectroscopy (MRS). A key result from \cite{ChenYC2024} is that \target\ is hosted by a powerful starburst galaxy with the instantaneous star-formation rate (SFR) exceeding $1,000\ M_{\odot}\textrm{ yr}^{-1}$, as traced by the bright \pah\ emission. This paper focuses on the analysis of the rest-frame optical emission lines observed with JWST/NIRSpec IFU. 

\begin{figure*}
    \begin{center}
    \begin{tabular}{cc}
    \includegraphics[width=0.51\textwidth]{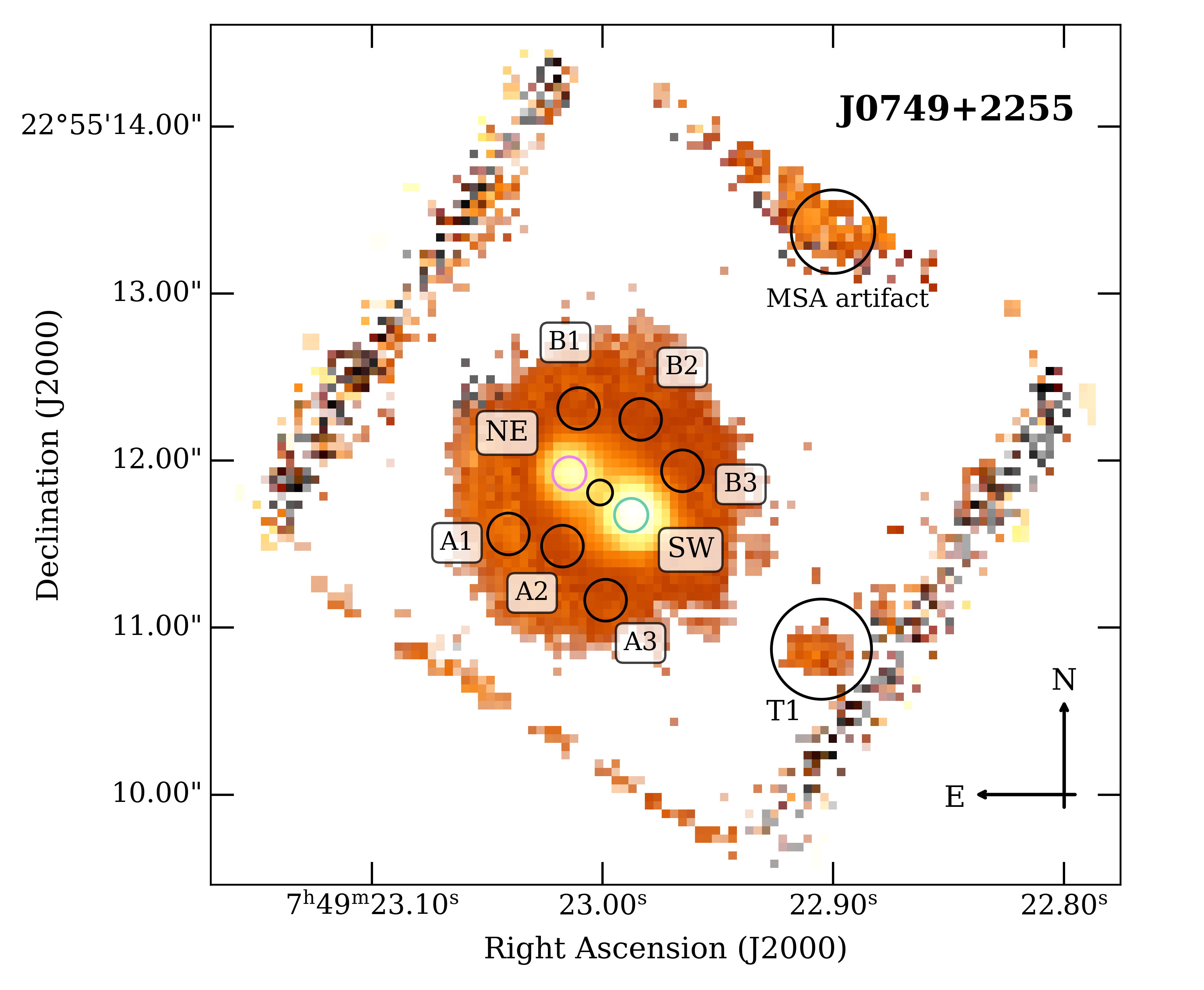} &
    \includegraphics[width=0.45\textwidth]{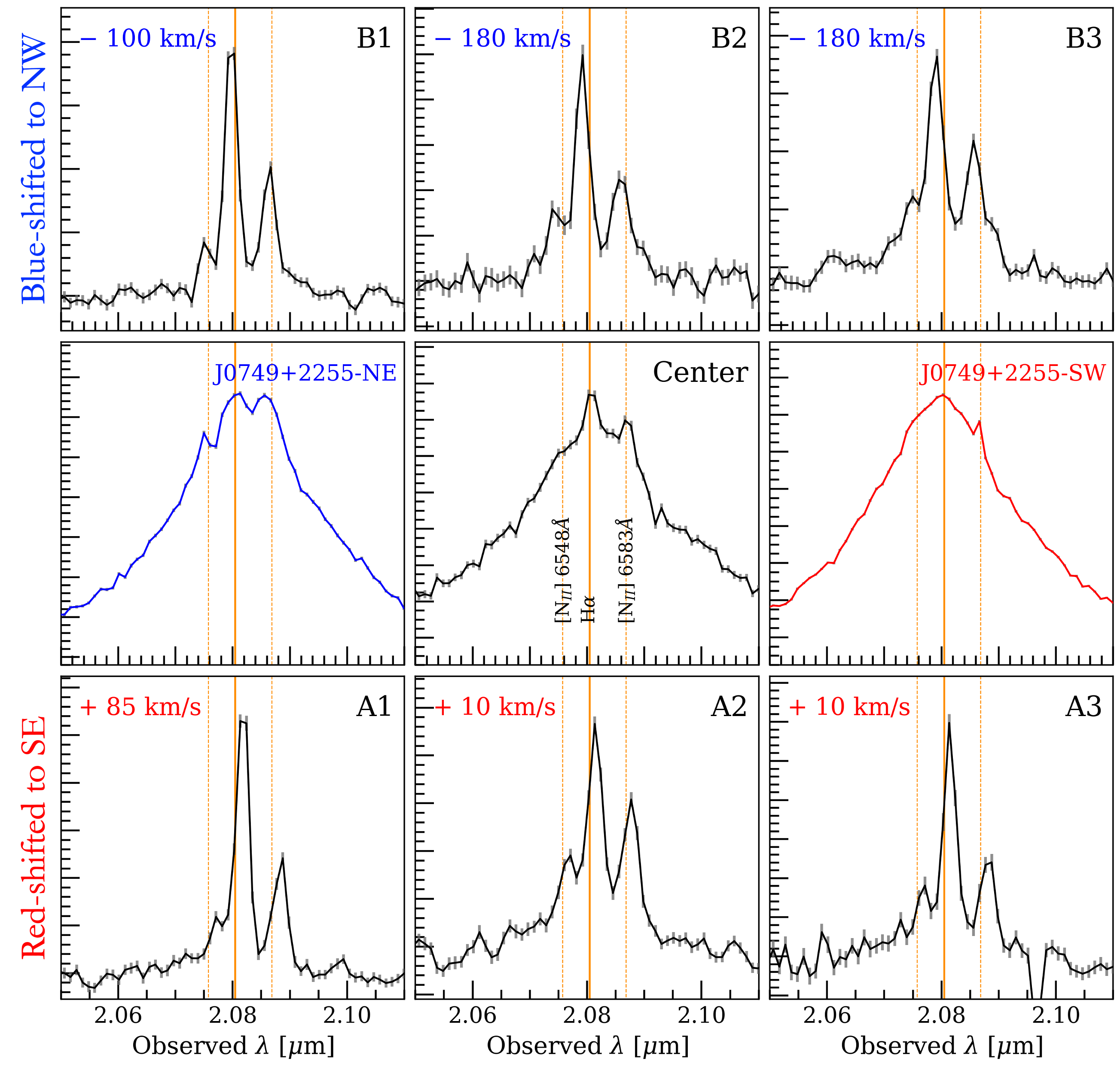}
    \end{tabular}
     \end{center}
    \caption{(Left) An integrated flux map of \target\ around the \ha+\nii\ line blend is shown in the log-scale. We apply a signal-to-noise (SNR) cut of $\textrm{SNR}\ >3$ to highlight the diffuse emission. Each circle overlay indicates the selected apertures for spectroscopic extraction, which is shown on the right panel. The red and blue colors correspond to \targSW\ and \targNE\, respectively, and the black circles indicate select apertures of the extended emission. There are significant noise artifacts along the detector edge illustrated in black blobs. We also indicate a major MSA artifact in the upper region of the detector. T1 marks a clear detection of a tidal feature or companion at the redshift of the quasar \citep{ChenYC2023a}. (Right) Each box corresponds to the different aperture spectra of the \ha+\nii\ line blend indicated on the left map. The fluxes are shown in the linear scale and have been normalized to highlight the line emission shape and centroids. The light gray errorbars indicate the flux uncertainty, which are nearly negligible. Two quasars, \targSW\ and \targNE, are surrounded by extended emission traced with extended narrow emission line regions with varying velocity offsets. The vertical orange lines indicate the \ha+\nii\ lines centered at $z=2.168$, the average systematic redshift.}
    \label{fig:totalMAP} 
\end{figure*}

\section{Observations and data reduction }\label{sec:dataRedux}
NIRSpec IFU observations were set up with two grating-filter combinations: G140M/F100LP and G235M/F170LP. This results in an effective wavelength coverage of $0.99-3.15$ \mum\ and a spectral resolution of $R\sim400-1000$ across the two gratings. There were no dedicated target verification exposures. We use the NRSIRS2RAPID readout mode for improved noise performance for an effective exposure time per integration of 380 sec. and a total exposure of 3282.498 sec. We use a 9-point dither pattern to improve the spatial sampling to accurately measure and characterize the PSF. The NIRSpec/IFU field-of-view (FOV) is $3\arcsec\times3\arcsec$, which corresponds to a physical scale of roughly $25\ \textrm{kpc}\times25\ \textrm{kpc}$. 

We reduce the NIRSpec data following the methods outlined by \cite{Vayner2023}. Data reduction was completed using the STScI JWST pipeline\footnote{\jwstpipurl} version 1.10.1 \citep{jwst1101} with CRDS version 11.16.2 and \texttt{jwst\_1077.pmap}. 

The first stage, \texttt{Detector1Pipeline}, performs standard infrared detector reductions, including dark subtraction, data quality flagging, bias subtraction, and cosmic ray removals, on uncalibrated files to produce rate files. We correct for $1/f$ noise \citep{Schlawin2020}. Then, these rate files are processed by the second stage, \texttt{Spec2Pipeline}, which assigns the world coordinate system to each frame, applies flat-field corrections, flux calibrates, and extracts the 2D spectra to build a 3D cube for each dither exposure. We use the \texttt{emsm} routine instead of the \texttt{drizzle} routine to build the 3D cubes\footnote{\jwstotherpipe}. \cite{Vayner2024} demonstrate that the \texttt{emsm} has a minimal detrimental effect on the PSF while improving the spectral ``wiggles'' that arise due to undersampling. This has been noted by other NIRSpec programs \citep[e.g.,][]{Wylezalek2022, Veilleux2023, Vayner2023}. We also skip the imprint subtraction step due to increased noise. After this step, we correct for flux contamination from the Micro-Shutter Assembly (MSA).

Finally, we combine the calibrated and corrected dither datacubes into a single datacube using in-house routines  \citep{Vayner2023, Veilleux2023}. They use the \python\ based \texttt{reproject} method to align dither exposures and combine them into a single datacube with a spatial resolution of $0.05\arcsec$ per spaxel. The final combined dithering pattern allows for a slightly larger FOV of roughly $4\arcsec \times 4\arcsec$, which is sufficient to capture any extended emission around the two quasars. Figure \ref{fig:totalMAP} shows a slice of the calibrated, MSA-corrected datacube centered around the observed \ha\ emission. We perform absolute flux calibration with the JWST/NIRSpec commissioning standard star observation of P330E. P330E is a G2V star that was observed with the same G140M/F100LP and G235M/F170LP IFU setups (PID: 1538). We reduced the standard star in the same way as the science cubes described above and flux calibrated with archival \hst/STIS spectra from the CALSPEC database \citep{Bohlin2014, Bohlin2015, Bohlin2022}. 

\begin{figure*}
	 \begin{center}
	 \begin{tabular}{c}
    \includegraphics[width=0.97\textwidth]{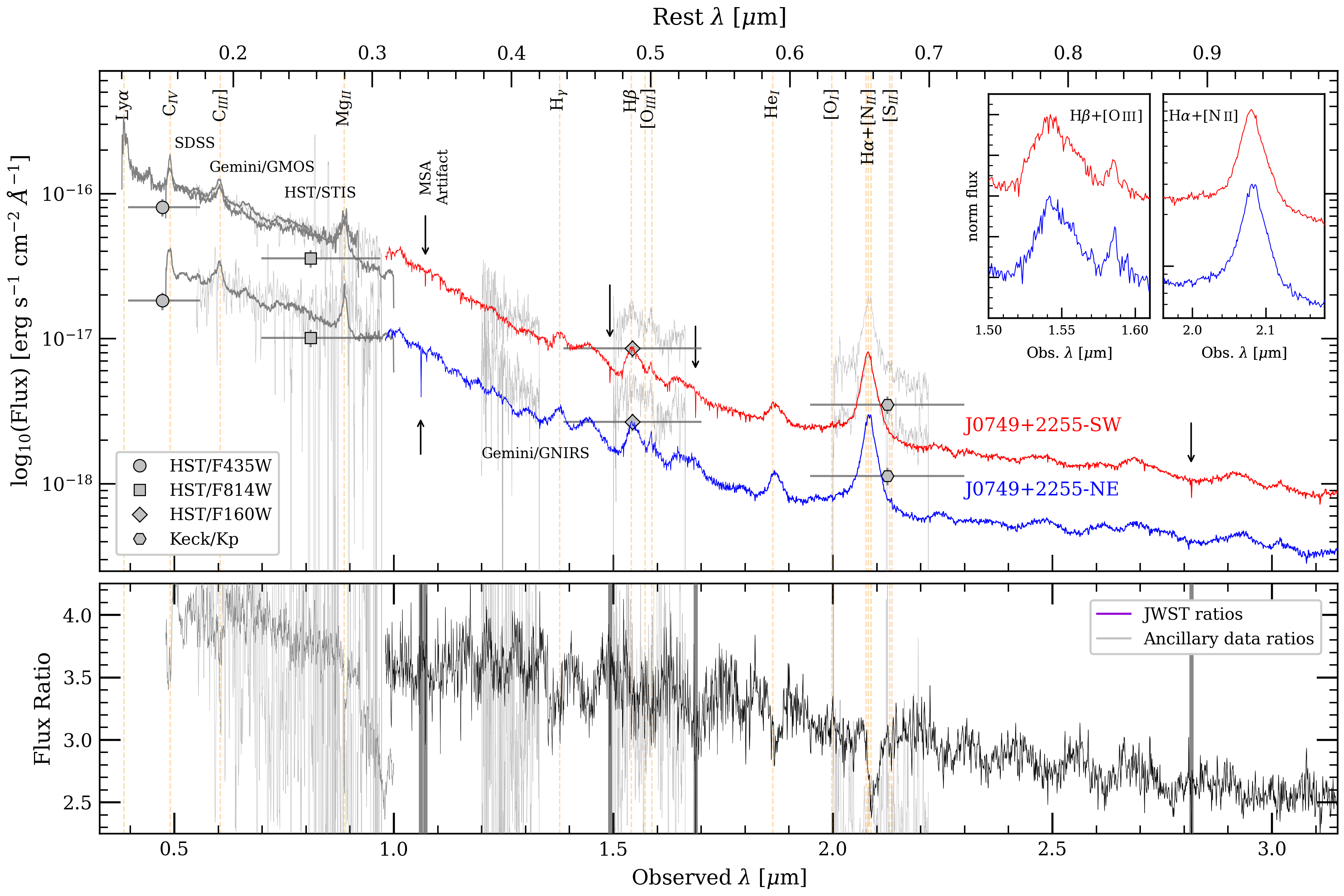}
	 \end{tabular}
	 \end{center}
	 \caption{(Top) Multi-wavelength observations of the two quasars in \target. We show data from \cite{ChenYC2023a}: unresolved SDSS spectroscopy in dark gray line, spatially-resolved slit spectroscopy (HST/STIS, Gemini/GMOS, and Gemini/GNIRS) in faint gray lines, and spatially-resolved HST/F475W/F814W/F160W and Keck/$K_p$ photometry are indicated with gray symbols. The spatially resolved JWST/NIRSpec aperture spectra are shown in red (\targSW) and blue (\targNE). The sharp line features that resemble absorption in the JWST spectra are artifacts due to imperfect MSA leakage correction; these regions are indicated with black arrows. The inset plots zoom into the \oiii+\hb\ and \ha+\nii\ regions with the fluxes normalized and offset for plotting purposes. (Bottom) The flux ratio of the two quasars (\fQratio): the JWST/NIRSpec data is in black and the ancillary spectra (HST and Gemini) are in light gray.} 
	 \label{fig:multiwave_qso_compare} 
\end{figure*}

After inspecting the science and ``leakcal'' exposure cubes, we discovered significant contamination from the MSA. This contamination appears as a bright, ``narrow line'' emission scattered throughout the final combined NIRSpec datacube in both spatial and spectral dimensions, affecting the regions of scientific interest. In each wavelength slice, the contaminants manifest as bright individual pixels or groups of pixels, spanning up to $0.8\arcsec$ in diameter. Each exposure contained over fifteen contaminated wavelength intervals. Fortunately, the MSA leakage appeared at consistent wavelengths and spatial positions across all dither exposures (both science and ``leakcal'' exposures) for each grating configuration. To mitigate it, we created a mask datacube that flagged bright pixels in the ``leakcal'' exposure with flux above the median background in each wavelength slice. The mask was then applied to every science dither exposure, effectively removing the MSA contamination. Unfortunately, this process resulted in a loss of signal in select pixels of the final datacube, which introduced non-astrophysical absorption-like features in the extracted aperture spectra. Figure \ref{fig:multiwave_qso_compare} shows aperture spectra taken from each quasar, with arrows marking the wavelength regions affected by the MSA ``absorption'' artifacts such as $\sim1.04\ \mu{m}$ and $\sim1.49\ \mu{m}$. 

\section{Spectral Analysis}\label{sec:specAnaly}
\subsection{Quasar properties}\label{sec:specAnaly:qso}
With JWST/NIRSpec IFU we confirm the detection of two quasars separated by $0.46\arcsec$ or $3.8\ \textrm{kpc}$. First, we examine the spectra of the two quasars by taking apertures with $r=2\textrm{ spaxels}$ ($0.1"$ or $\sim0.8\textrm{ kpc}$) radii. In Figure \ref{fig:multiwave_qso_compare} we plot the extracted JWST spectra and compare them with the previous observations from \cite{ChenYC2022}. We find that the quasar to the southwest (\targSW) is on average about $\times3$ brighter at the observed $\sim2$ \mum\ ($\sim0.63$ \mum, rest-frame) than the quasar to the northeast (\targNE) shown in Figure \ref{fig:totalMAP}. Both quasars exhibit unobscured, Type 1 quasar features like the blue continuum and broad Balmer emission lines. 

\begin{figure*}
	 \begin{center}
	 \begin{tabular}{c}
    \includegraphics[width=0.85\textwidth]{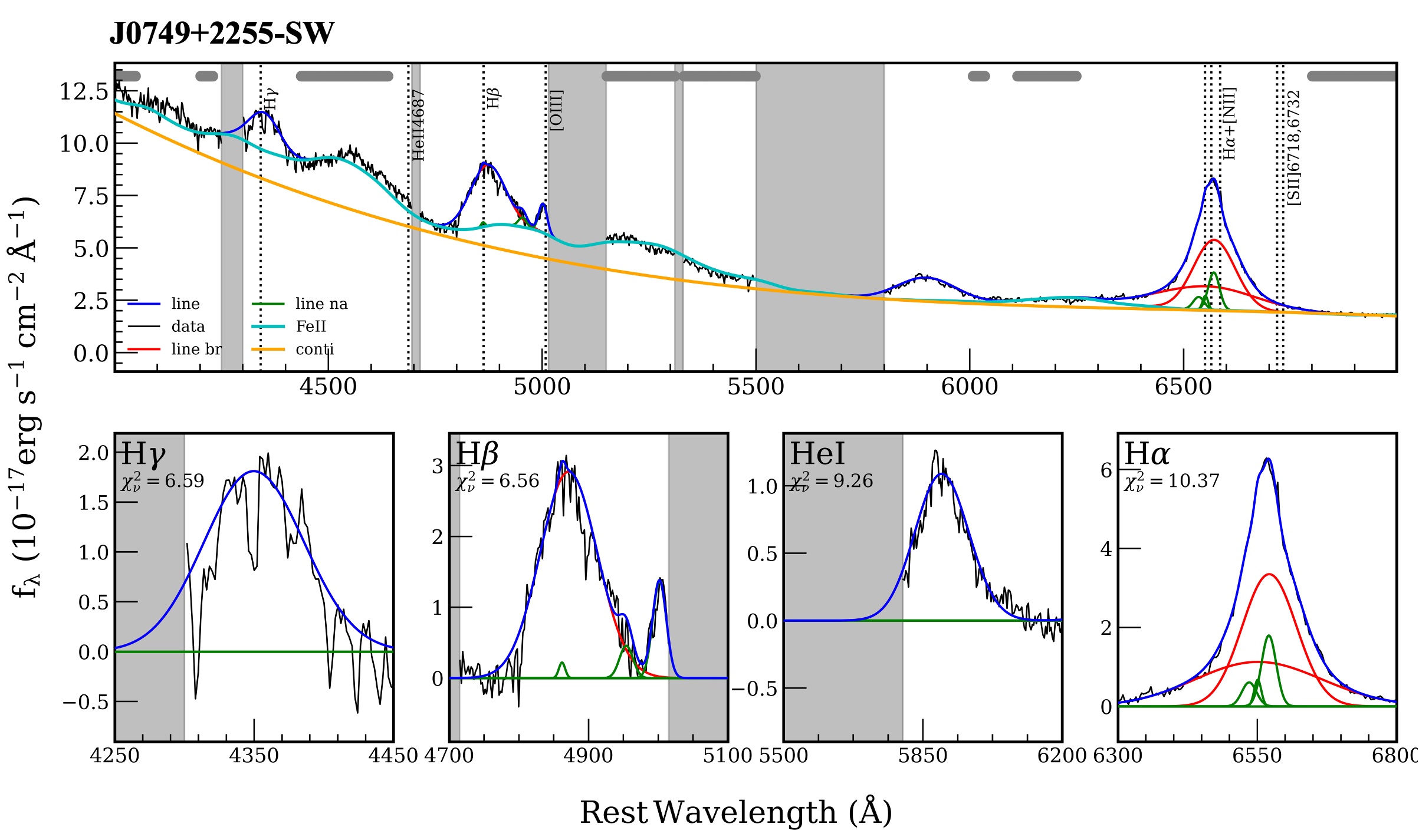} \\
    \includegraphics[width=0.85\textwidth]{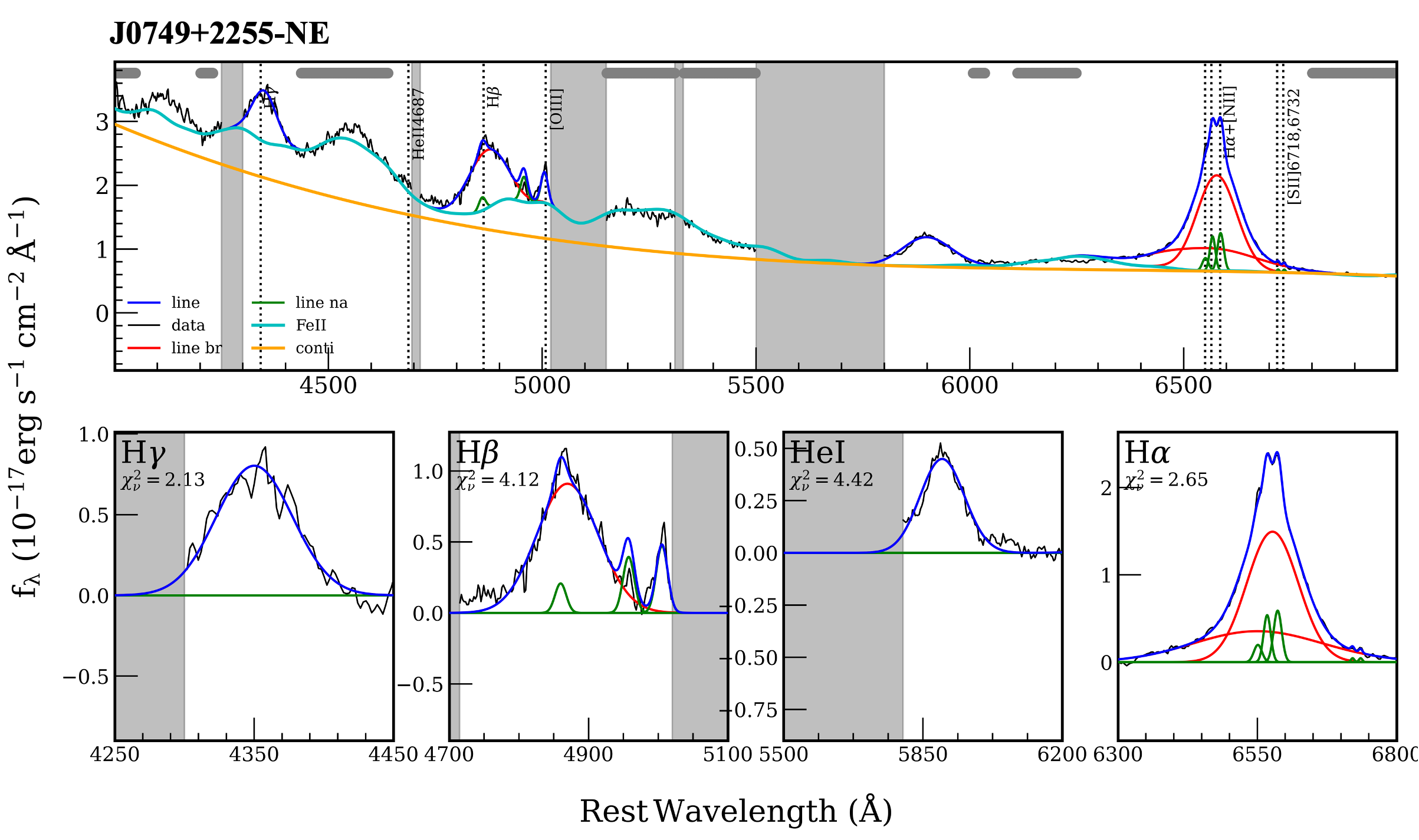} \\
	 \end{tabular}
	 \end{center}
	 \caption{The \pyqsofit\ results for the brighter \targSW\ (top) and fainter \targNE\ (bottom) quasars. The shaded regions in the \targSW\ spectrum indicate the masked regions due to the MSA light leakage. The dark gray bars at the top of the plot indicate the wavelength regions used to fit the continuum and the \feii\ lines. We also show select emission line fits of \ha, \nii, \hei, \hb, \oiii, and \hg.} 
	 \label{fig:pyqsofit} 
\end{figure*}

Interestingly, we find that the two quasars have similar, yet subtly different spectra. If we take the flux ratio of the two quasars (\fQratio), we find that their continua and emission lines are not a one-to-one match. The \fQratio\ ratio has a blue slope such that \targSW\ is $\times3.5$ brighter at $\sim1$ \mum\ and $\times3$ brighter at $\sim2.5$ \mum\ in the observed frame, as shown in Figure \ref{fig:multiwave_qso_compare}. This means that \targNE\ appears slightly redder than \targSW. We can also see subtle differences in the emission line profile shapes (e.g. \ha+\nii, \hb+\oiii, and \hei) between the two quasars. In the following sections, we fit the nuclear spectra and discuss the emission line properties in detail.

\subsubsection{Fitting the quasar spectra}
We use \pyqsofit \footnote{\qsofiturl} \citep{GuoH2018,Shen2019a} to fit the 1-D spectra of the two quasars to estimate the systemic redshift and to measure the emission-line properties of each quasar. Following the treatment in \cite{ChenYC2023a}, we model each quasar spectrum as a linear combination of a pseudo-continuum (power-law, polynomial, and \feii\ components), broad emission lines, and narrow emission lines. We fit for the following emission lines: Balmer lines (\ha, \hb, and \hg), \hei\ $\lambda5877$\AA, \nii$\lambda\lambda6549$\AA,$6585$\AA, \sii$\lambda\lambda 6718$\AA,$ 6732$\AA, and \oiii$\lambda\lambda 4959$\AA,$ 5007$\AA\ with a combination of broad and narrow components. We fit over the rest-frame wavelengths $4000-7000$\AA\ ($1.2-2.2$ \mum\ in the observed frame). We freely fit for the line centroids, line widths, and amplitude. The redshift prior is set to $z=2.168$, and shifts to the line centroids are used to refine the systemic redshifts. We summarize the \pyqsofit\ fit results in Figure \ref{fig:pyqsofit} and Table \ref{tab:fitproperties}.

\begin{table*}
\centering
\caption{The \pyqsofit\ quasar line fit results of the brighter \targSW\ (brighter) and fainter \targNE. The calculated SMBH accretion properties corresponding to these fit values are shown in Table \ref{tab:qsoprop}. The systemic redshift is calculated using the \oiii$\lambda5007$\AA\ narrow line emission. We show the calculated $\log_{10}L_{bol}$ values, which assume $\textrm{BC}_{5100}=4.3\pm1.3$ \citep{Krawczyk2013}. We also show $\log_{10}L_{bol}$ values for $\textrm{BC}_{5100}=7.8\pm1.7$ \citep{Krawczyk2013, Richards2006b} in parenthesis, which results in larger \Lbol\ values. All errors quoted here are the $1\sigma$ uncertainties derived from \pyqsofit\ results, which account for the errors in the input spectrum, and the continuum uncertainty due to \feii\ contamination.  } 
\label{tab:fitproperties} 
\begin{tabular}{lclll}
    \hline
    Nuclear emission-lines & Broad/Narrow & Units & \targSW & \targNE \\ 
    \hline
    $z_{sys, [\textrm{O}~\textsc{iii}]}$  & - & - & $2.1670 \pm 0.0001$ & $2.1677 \pm 0.0006$ \\
    \ha\ Flux & Broad 1 & (\ergflux) & 
    $(3.5\pm0.5)\times10^{-15}$ & $(1.6 \pm 0.3)\times10^{-16}$ \\
    \ha\ Luminosity & Broad 1 & (\ergs) & 
    $(1.3 \pm 0.2)\times10^{44}$ & $(5.8 \pm 0.4)\times10^{43}$ \\
    \ha\ FWHM & Broad 1 & (\kms) & 
    $5000\pm 500$ & $4900\pm 500$ \\
    \ha\ Flux & Broad 2 & (\ergflux) & 
    $(3.0\pm0.4)\times10^{-15}$ & $(1.0 \pm 0.1)\times10^{-15}$ \\
    \ha\ Luminosity & Broad 2 & (\ergs) & 
    $(1.1 \pm 0.1)\times10^{44}$ & $(3.6 \pm 0.5)\times10^{43}$ \\
    \ha\ FWHM & Broad 2 & (\kms) & 
    $12,300\pm 500$ & $12,500\pm 500$ \\
    \ha\ Flux & Narrow & (\ergflux) & 
    $(1.1 \pm 1)\times10^{-16}$ & $(2.0 \pm 0.3)\times10^{-16}$ \\
    \ha\ Luminosity & Narrow & (\ergs) & 
    $(6 \pm 4)\times10^{42}$ & $(8.5 \pm 0.5)\times10^{42}$ \\
    \ha\ FWHM & Narrow & (\kms) & 
    $682\pm 500$ & $830\pm 500$ \\
    \hb\ Flux & Broad & (\ergflux) & 
    $(3.0 \pm 0.4)\times10^{-15}$ & $(1.0 \pm 0.5)\times10^{-15}$ \\
    \hb\ Luminosity & Broad & (\ergs) & 
    $(1.1 \pm 0.1)\times10^{44}$ & $(4 \pm 1)\times10^{43}$ \\
    \hb\ FWHM & Broad & (\kms) & 
    $5900\pm 700$ & $6600\pm 700$ \\
    \hb\ Flux & Narrow & (\ergflux) & 
    $(5 \pm 10)\times10^{-17}$ & $(4 \pm 2)\times10^{-17}$ \\
    \hb\ Luminosity & Narrow & (\ergs) & 
    $(2 \pm 4)\times10^{42}$ & $(1 \pm 2)\times10^{42}$ \\
    \hb\ FWHM & Narrow & (\kms) & 
    $658\pm 500$ & $1400\pm 500$ \\
    \hg\ Flux & Broad & (\ergflux) & 
    $(1.4 \pm 0.3)\times10^{-15}$ & $(9 \pm 1)\times10^{-15}$ \\
    \hg\ Luminosity & Broad & (\ergs) & 
    $(5.0 \pm 1.1)\times10^{43}$ & $(3.3 \pm 1)\times10^{43}$ \\
    \hei\ Flux & Broad & (\ergflux) & 
    $(1.9 \pm 0.1)\times10^{-15}$ & $(2.0 \pm 0.7)\times10^{-15}$ \\
    \hei\ Luminosity & Broad & (\ergs) & 
    $(6.6 \pm 4.6)\times10^{43}$ & $(7.0 \pm 0.6)\times10^{43}$ \\
    \oiii\,$\lambda5007$\AA\ Flux & Narrow & (\ergflux) & 
    $(3 \pm 1)\times10^{-16}$ & $(3.2 \pm 0.4)\times10^{-16}$ \\
    \oiii\,$\lambda5007$\AA\ Luminosity & Narrow & (\ergs) & 
    $(1.0 \pm 0.5)\times10^{43}$ & $(1.1\pm 0.1)\times10^{43}$ \\
    \oiii\,$\lambda5007$\AA\ FWHM & Narrow & (\kms) & 
    $1200\pm 500$ & $1400\pm 500$ \\
    \feii\ optical FWHM  & Broad & (\kms) & $6200\pm 550$ & $4900\pm 550$ \\
    $\log_{10}(L_{5100}/\textrm{erg s}^{-1})$ & - & - &$45.87^{+0.07}_{-0.10}$ & $45.28^{+0.07}_{-0.10}$ \\
    $\log_{10}(L_{bol}/\textrm{erg s}^{-1})$ & - & - &$46.49^{+0.13}_{-0.19}$ ($46.75^{+0.13}_{-0.19}$) & 
    $45.91^{+0.14}_{-0.20}$ ($46.17^{+0.13}_{-0.19}$)\\
    \hline
\end{tabular}
\end{table*}
\begin{table*}
\centering
\caption{The SMBH properties calculated based on the broad component Balmer line fits based on \pyqsofit\ in Table \ref{tab:fitproperties}. The $(a,b,c)$ coefficients for the \Mbh\ calculations use the calibration by \cite{Shen2012} and \cite{Shen2024}. We quote their mean scatter in \Mbh\ for the different calibrations. The mean $\log_{10}(M_{BH}/M_{\odot})$ are $9.31\pm0.22$ and $9.04\pm0.23$ for \targSW\ and \targNE, respectively. We calculate two values of \eddrat, assuming different bolometric correction for $\textrm{BC}_{5100}=4.3\pm1.3$ \citep{Krawczyk2013} and for $\textrm{BC}_{5100}=7.8\pm1.7$ \citep{Krawczyk2013, Richards2006b} in parenthesis. The mean \eddrat\ values for \targSW\ and \targNE, respectively, are $0.14\ (0.25)\pm0.2$ and $0.06\ (0.11)\pm0.1$.} 
\label{tab:qsoprop} 
\begin{tabular}{lcccc}
    \hline
    Method & $(a,b,c)$ & Measurement & 
    \targSW & \targNE \\ 
    \hline
    ${\rm FWHM}(\rm H \alpha)+L_{5100}$
    & $(1.390, 0.555, 1.873)$
    & $\log_{10}(M_{BH}/M_{\odot})$ 
    & $9.40 \pm 0.18$  &  $9.03 \pm 0.20$ \\
    & & \eddrat\ & 
    $0.10\ (0.19) \pm 0.09$ & $0.05\ (0.11) \pm 0.11$ \\
    
    ${\rm FWHM}(\rm H \alpha)+L_{\rm H \alpha}$
    & $(2.216, 0.564, 1.821)$
    & $\log_{10}(M_{BH}/M_{\odot})$ 
    & $9.08 \pm 0.14$  &  $8.80 \pm 0.16$ \\
    & & \eddrat\ & 
    $0.21\ (0.39) \pm 0.07$ & $0.06\ (0.18) \pm 0.10$ \\

    ${\rm FWHM}(\rm H \beta)+L_{5100}$
    & $(0.85, 0.5, 2.0)$
    & $\log_{10}(M_{BH}/M_{\odot})$ 
    & $9.34 \pm 0.37$  &  $9.01 \pm 0.37$ \\
    & & \eddrat\ & 
    $0.12 \ (0.21) \pm 0.17$ & $0.02\ (0.09) \pm 0.18$ \\
    
    ${\rm FWHM}(\rm H \beta)+L_{\rm H \beta}$ 
    & $(1.963, 0.401, 1.959)$
    & $\log_{10}(M_{BH}/M_{\odot})$ 
    & $9.37 \pm 0.12$  &  $9.30 \pm 0.12$ \\
    & & \eddrat\ & 
    $0.11\ (0.20) \pm 0.06$ & $0.04\ (0.07) \pm 0.09$ \\

    \hline
\end{tabular}
\end{table*}

As hinted from the rest-frame UV spectra \citep{ChenYC2023a}, \target\ displays strong and broad optical \feii\ emission with FWHM reaching nearly $\sim6000\ \textrm{km s}^{-1}$. We detect two broad \ha\ line components with FWHM $\sim5000\ \textrm{km s}^{-1}$ and $\sim10,000\ \textrm{km s}^{-1}$. In addition to the strong \feii\ emission, both quasars exhibit weak \oiii\ emission compared to \hb. This resulted in contamination in the \hb$+$\oiii\ line blend, so it was unclear whether the broadest components of \hb\ and \oiii\ are real or artifacts from insufficient (or even oversubtraction) of the \feii. Despite the best \pyqsofit\ continuum decomposition efforts, the likely and conservative \feii\ contamination is about 10-20\%, which dominates the error. 

There are subtle differences in the emission line profiles in contrast to their similar appearances in their optical continuum properties. While the FWHM of the broad component of \ha\ between \targSW\ and \targNE\ are similar to within error, the narrow line components around \hb+\oiii\ and \ha+\nii\ differ. The narrow emission lines associated with \targNE\ are more prominent, resulting in a ``pointier'' spectrum than those of \targSW. We find that, despite contamination from \feii, the equivalent width of the \oiii\ emission from \targNE\ is greater than that of \targSW. 

Despite the caveat of significant \feii\ contamination, we measure the systemic redshift of each quasar based on the centroid shift of the narrow \oiii$\lambda5007$\AA\ emission line. For completeness, we also compare the centroid shifts of the narrow \ha\ and \nii\ emission lines. While the \oiii-based centroid measurements are consistent for both quasars, the \ha\ and \nii\ based centroid shift for \targNE\ differ, so we take the mean value.  The mean centroid shifts from $z=2.168$ for \targSW\ and \targNE\ are $6\pm0.5$\AA\ and $3\pm1$\AA, which corresponds to systemic redshifts: $\langle z_{SW}\rangle = 2.1665\pm0.0001$ for \targSW\ and $\langle z_{NE}\rangle = 2.16805\pm0.0006$ for \targNE, which differ previous results \citep{ChenYC2023a}, likely due to improved sensitivity. 

\subsubsection{Quasar luminosity}
From the best-fit \feii-subtracted continuum model, we determine the monochromatic optical continuum at $\lambda=5100$\AA, $L_{5100}$, which we use to calculate the bolometric luminosity, $L_{bol}$, and the black hole mass, \Mbh. Both quasars are optically luminous. The uncertainty in $L_{5100}$ is derived from the $1\sigma$ \pyqsofit\ error and the \feii\ contamination error. The measured $L_{5100}$ values are $(1.9\pm0.4)\times10^{45}\ \textrm{erg s}^{-1}$ for \targNE\ and $(7.2\pm1.5)\times10^{45}\ \textrm{erg s}^{-1}$ for \targSW. 

We calculate \Lbol\ using the relation, $L_{bol} = \lambda L_{\lambda} \times \textrm{BC}_{\lambda}$, where $\textrm{BC}_{\lambda}$ is the wavelength-dependent bolometric correction. To convert from $L_{5100}$, we use $\textrm{BC}_{5100}=4.3\pm1.3$ \citep{Krawczyk2013}, which is calibrated in the wavelength range 30 \mum\ to 2 keV that avoids IR and X-ray double-counting. We propagate the errors from $L_{5100}$ and $\textrm{BC}_{\lambda}$ to calculate the uncertainty in \Lbol. We obtain \Lbol\ values: $(8.2\pm3.0)\times10^{45}\ \textrm{erg s}^{-1}$ for \targNE\ and $(3.1\pm1.1)\times10^{46}\ \textrm{erg s}^{-1}$ for \targSW. The calculated \Lbol\ of the two quasars are distinct to only $1\sigma$ with a small difference of a few factors. It is rather interesting to find a quasar pair with similar emission properties. We also calculate \Lbol\ assuming $\textrm{BC}_{5100}=7.9\pm1.7$ \citep{Krawczyk2013,Richards2006b}.

\subsubsection{SMBH properties}
Using the best-fit results, we calculate the single-epoch virial black hole masses, \Mbh, with the \cite{Shen2012} formalism based on reverberation mapping:
\begin{equation}
    \begin{split}
        \log_{10}\bigg(\frac{M_{BH}}{M_{\odot}}\bigg) = a+  b \log_{10}\bigg(\frac{\lambda L_{\lambda} }{10^{44} \textrm{ erg s}^{-1}}\bigg) \\
         + c  \log_{10}\bigg(\frac{ v_{\textrm{FWHM}}}{  \textrm{km s}^{-1}    }\bigg) 
    \end{split}
\label{eq:BHvir}
\end{equation}

We calculate \Mbh\ using two different calibrations: (a) $L_{5100}$ continuum and the broad \ha\ or \hb\ line FWHM \citep{Shen2012, Shen2024}, and (b) the broad \ha\ or \hb\ line luminosity and FWHM \citep{Shen2012}, where $a$, $b$, and $c$ correspond to different calibration coefficients. The average minimum uncertainty of \Mbh\ based on the calibrations is 0.2 dex. We propagate the measured uncertainty in $L_{5100}$ and the \Mbh\ calibration to calculate the total uncertainty in \Mbh. Lastly, we combine the \Mbh\ and \Lbol\ estimates to calculate the Eddington luminosity, $L_{Edd}$, and the Eddington ratio, $\lambda_{Edd}=L_{bol}/L_{Edd}$. We propagate the error in the calculations $L_{5100}$ and \Mbh accordingly to determine the uncertainty in \eddrat. The calibration coefficients for the Eq.\ref{eq:BHvir}, \Mbh, \Ledd, and \eddrat\ calculations are listed in Table \ref{tab:qsoprop}. We obtain consistent \Mbh\ estimates using the different calibrations. The average \Mbh\ estimates for the different methods are $10^{9.31\pm0.22}\ \textrm{M}_{\odot}$ and $10^{9.04\pm0.23}\ \textrm{M}_{\odot}$ for \targSW\ and \targNE, respectively. Assuming $\textrm{BC}_{5100}=4.3\pm1.3$, we obtain  $\lambda_{Edd}=0.14\pm0.2$ and  $0.06\pm0.09$ for \targSW\ and \targNE, respectively. We obtain larger \eddrat\ estimates using $\textrm{BC}_{5100}=7.9\pm1.7$ with $\lambda_{Edd}=0.25\pm0.2$ and $0.11\pm0.1$, respectively, which is similar to previous estimates \citep{ChenYC2023a}. Although the choice of  $\textrm{BC}_{5100}$ affects the \Lbol\ and \eddrat\ calculations, the uncertainty in the continuum dominates the error in \Mbh, \Lbol, and \eddrat. Effectively the two \eddrat\ estimates for each quasar are consistent within error. Interestingly, both quasars have similar SMBH masses and Eddington ratios.

\subsubsection{Velocity offsets}
Based on the observations of two quasars with similar spectral (continuum and emission lines) and accretion properties, it would be natural to assume that \target\ is a gravitationally lensed pair rather than a physical dual quasar. Although \pyqsofit\ results indicate differences in the spectrum (line centroids and shape), which suggest a dual quasar, even gravitational lensing may produce images with slightly different spectra due the temporal variations in the inherent nuclear emission. One definitive test is to examine the velocity offsets. If the two quasars are indeed a physical pair, then we expect velocity offsets in the emission lines associated with each quasar. We investigate this using two methods: (a) cross-correlating the two spectra; and (b) comparing the centroids of the narrow emission lines based on the \pyqsofit\ results. 

We cross-correlate the \hb+\oiii\ and \ha+\nii\ emission lines systems and use the flux ratio between the two quasars (\fQratio) as a proxy. We fix the spectrum of \targSW, then incrementally offset the spectrum of \targNE\ in highly sampled steps, $\Delta\lambda$, and calculate the root-mean-square (RMS) spread in \fQratio\ as a function of $\lambda$. The total velocity offset, $\Delta v$, is the total shift that best minimizes $\textrm{RMS}(f_{\rm SW}/f_{\rm NE},\lambda)$ by fitting a smooth power-law. In Figure \ref{fig:multiwave_qso_compare} we see large deviations in the observed \fQratio\ around the emission lines, likely due to the velocity offsets between the two quasars. At the calculated $\Delta v$, those \fQratio\ deviations disappear. To calculate the uncertainty in the measured $\Delta v$, we perform a Monte Carlo simulation of the cross-correlation between two velocity-shifted spectra with varying spectral properties (i.e.~line amplitudes, line width, noise levels, velocity offset, and spectral sampling). We find a mean $\Delta v$ recovery of $10\pm5\ \textrm{km s}^{-1}$ (measured vs.~true value). This high recovery rate is likely attributed to comparing the line shape rather than the line centroids, usually determined by fitting Gaussian(s). From the simulation, we find that the key driver of uncertainty is the wavelength sampling of the spectrum; in fact, the method performs best when the line shapes of the two spectra are similar as in the case of \targSW\ vs.~\targNE. We measure a total velocity offset between the \ha+\nii\ lines of $\Delta\lambda=$5.2\AA\ or $\Delta v=240\pm15\ \textrm{km s}^{-1}$ with at least $3\sigma$ confidence.

Similarly, we compare the centroids of the narrow \oiii\ emission lines based on \pyqsofit. Although \oiii\ is fainter than the Balmer lines, the \oiii\ centroids are more reliable since the Balmer lines show complex kinematics requiring multiple broad components. We measure a \oiii\ centroid shift of $\Delta\lambda \approx3.5$\AA, which corresponds to $\Delta v\sim239 \pm 100$ \kms\ with only $<2\sigma$ confidence. The measured $\Delta v$ is consistent with the cross-correlation results, despite the greater uncertainty. Although these two tests indicate the presence of a velocity offset between the two quasars, the results are not definitive due to the large uncertainties. 

\begin{figure*}
	 \begin{center}
  \begin{tabular}{ccc}
   \includegraphics[trim={0.3cm 0.3cm 0.4cm 0cm},clip,width=0.32\textwidth]{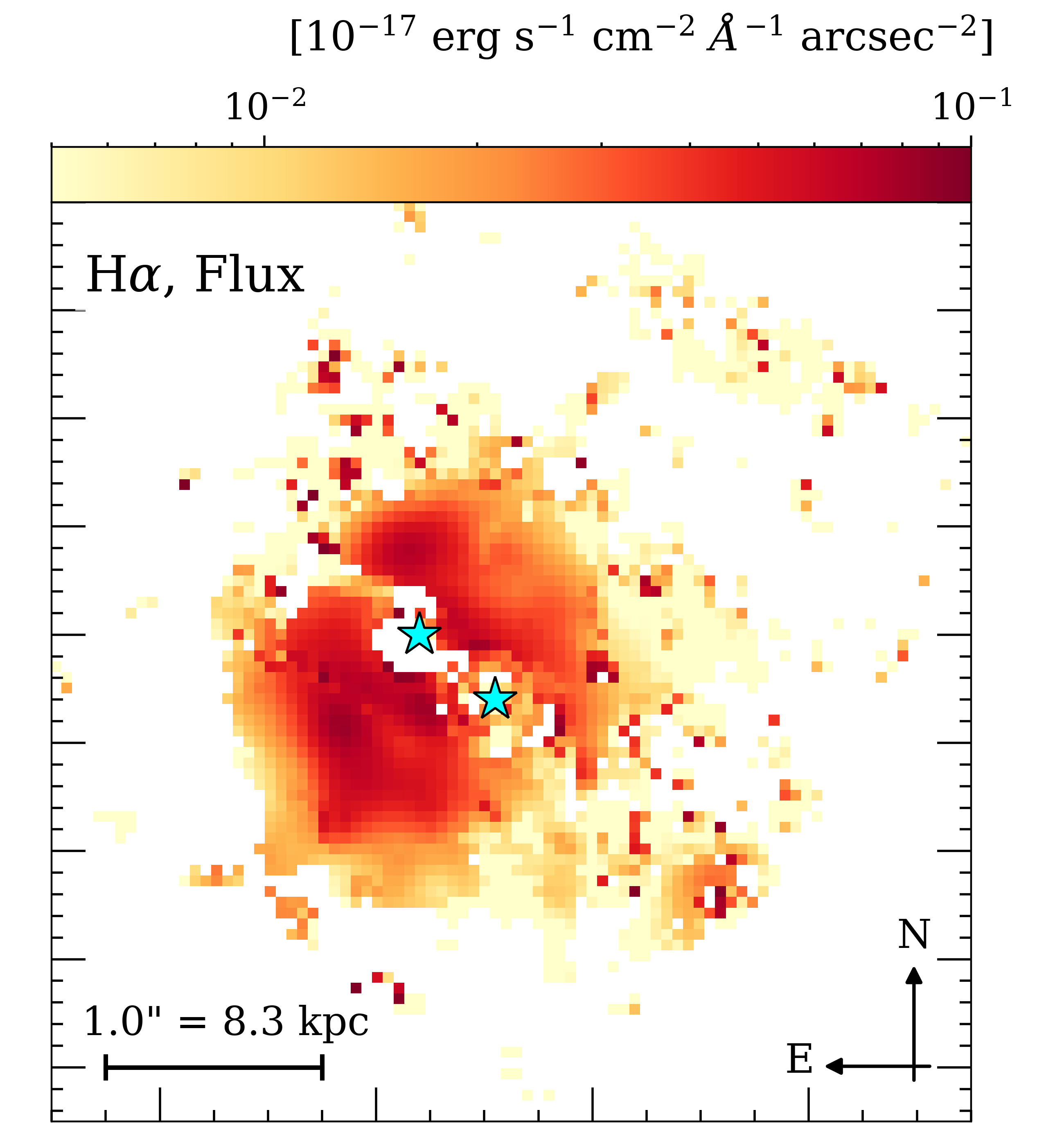}&
   \includegraphics[trim={0.3cm 0.3cm 0.4cm 0cm},clip,width=0.32\textwidth]{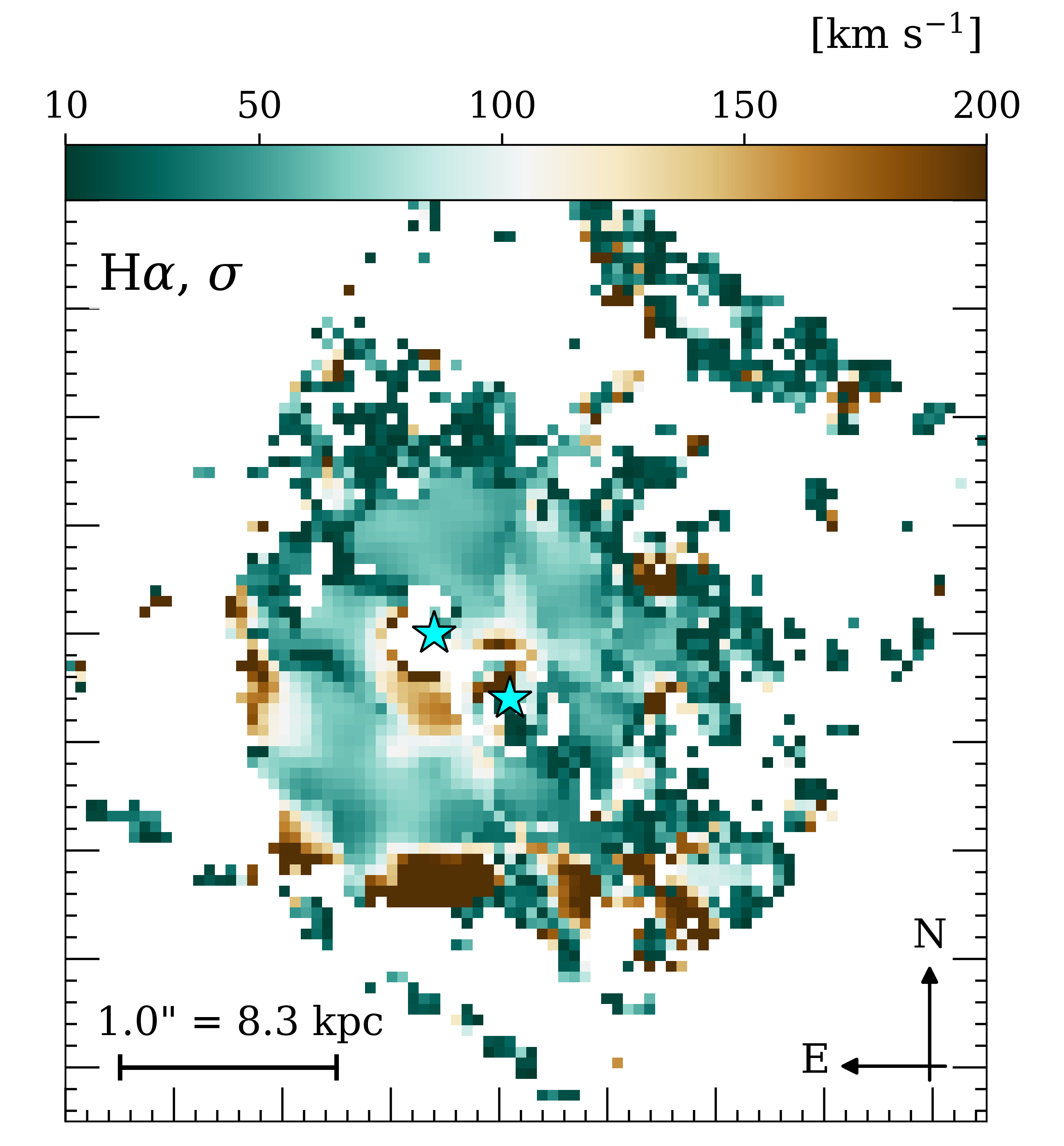}&
   \includegraphics[trim={0.3cm 0.3cm 0.4cm 0cm},clip,width=0.32\textwidth]{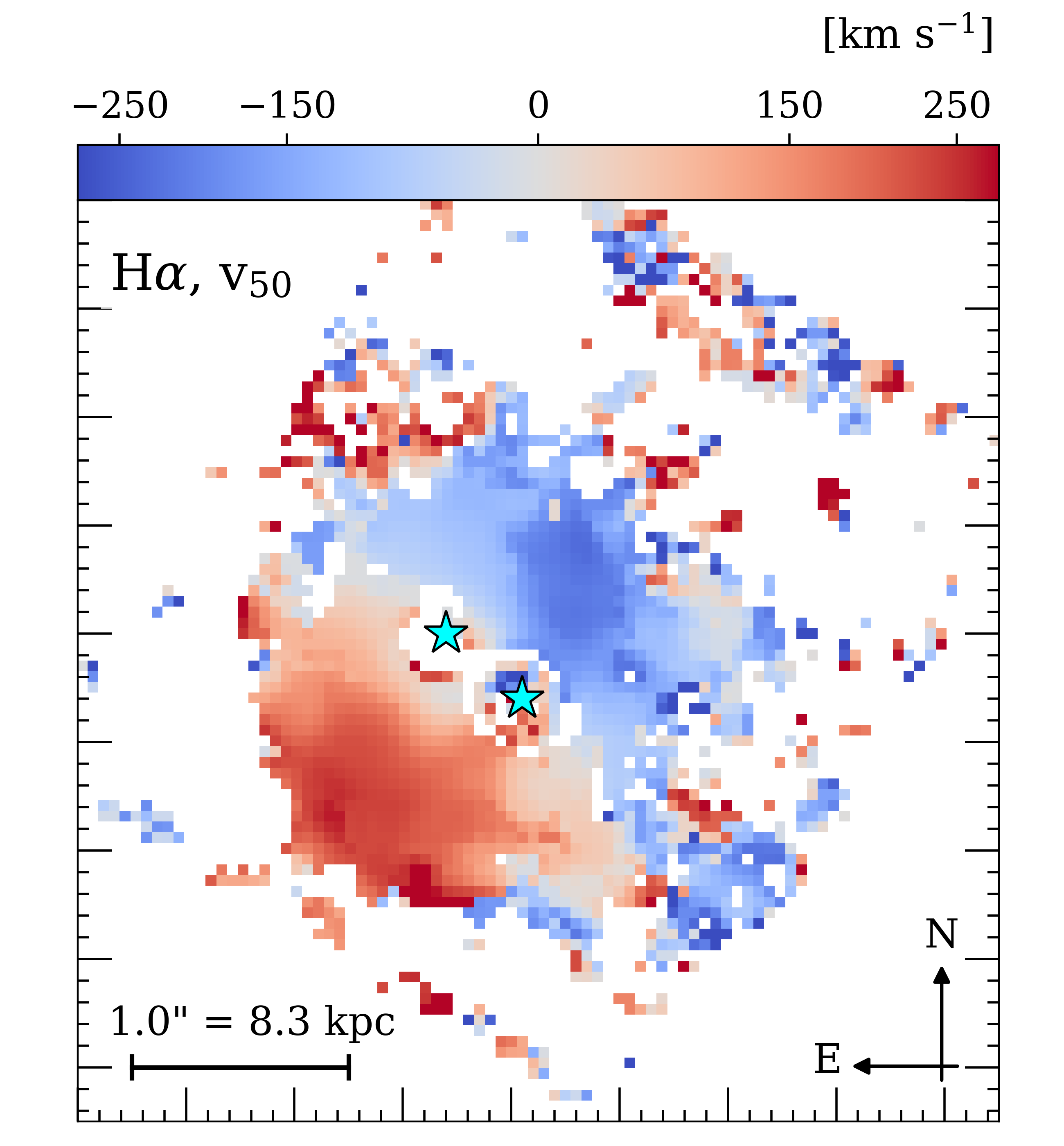}
	 \end{tabular}
	 \end{center}
	 \caption{Intensity (left), velocity dispersion (center), velocity (right) maps of the quasar-subtracted, narrow-line \ha\ produced with \qdfit. Each cyan star represents the location of the subtracted quasars: \targNE\ and \targSW. We see that the kinematic rotation is misaligned by $90^{\circ}$ with respect to the two quasars. } 
	 \label{fig:map_ha} 
\end{figure*}

\subsection{Quasar host galaxy properties}\label{sec:specAnaly:host}
In Figure \ref{fig:totalMAP} we see indications of extended narrow line \ha\ emission surrounding the two quasars. The \ha\ emission is roughly at the same redshift with indications of red-/blue-shifted velocity shifts with respect to the quasars. This is a clear detection of the host galaxy(ies) of the two quasars in \target. We describe our method of extracting the faint host galaxy emission around the quasars.

\subsubsection{PSF subtraction and line fitting of diffuse emission}

A major challenge to studying the faint emission around a quasar is that the quasar typically outshines its host galaxy by tens or hundreds of times. Studying the host galaxy usually involves the challenging task of carefully modeling and removing the quasar emission to extract the faint extended emission of the host galaxy. This problem is further compounded in \target\ since we have two bright quasars to account for. 

We use \qdfit\footnote{\qdfiturl} to model and subtract the unresolved point source emission of the quasar PSF to reveal the faint extended emission \citep{q3dfit2023}. \qdfit\ performs maximal-contrast subtraction of the quasar spectral PSF by using the spectral differences between the quasars and their host galaxies. \qdfit\ simultaneously fits the quasar emission, based on an empirically determined quasar spectrum, and the host galaxy emission model consisting of a 2-order polynomial continuum and emission lines across the NIRSpec datacube \citep{Vayner2023}. The quasar template is set by extracting an aperture spectrum centered on the quasar: either the brightest spaxel (default setting) or the user-defined spaxel. Interestingly, although \qdfit\ fits the unresolved quasar component, the reconstructed quasar emission matches the quasar PSF, as demonstrated by \cite{Vayner2024} and \cite{LiuW2024}. This means that \qdfit\ successfully models and subtracts the quasar PSF emission.

\target\ is a system of two quasars with two distinct spectra, so we require the removal of two PSFs with two separate quasar templates. Unfortunately, \qdfit, out of the box, is only optimized for a galaxy system hosting one quasar. Since it is possible to set the quasar template on a user-defined spaxel, we run the quasar decomposition fits twice; one decomposition uses the quasar template centered on \targSW\ and another uses the quasar template centered on \targNE. Each of the quasar spectra is extracted using a circular aperture with a 2-spaxel radius ($0.1\arcsec$). Unlike seeing-limited, ground-based observations, JWST in the infrared has a wavelength-dependent PSF. To account for the wavelength dependence, \qdfit\ applies a wavelength-dependent correction factor. This method is sufficient when fitting the emission line features. We make no assumptions about the shape of the host galaxy emission. Using this method, we effectively produce two separate PSF-subtracted datacubes corresponding to \targSW\ and \targNE. 

\begin{figure*}
	 \begin{center}
  \begin{tabular}{ccc}
    \includegraphics[trim={0.3cm 0.3cm 0.4cm 0cm},clip,width=0.32\textwidth]{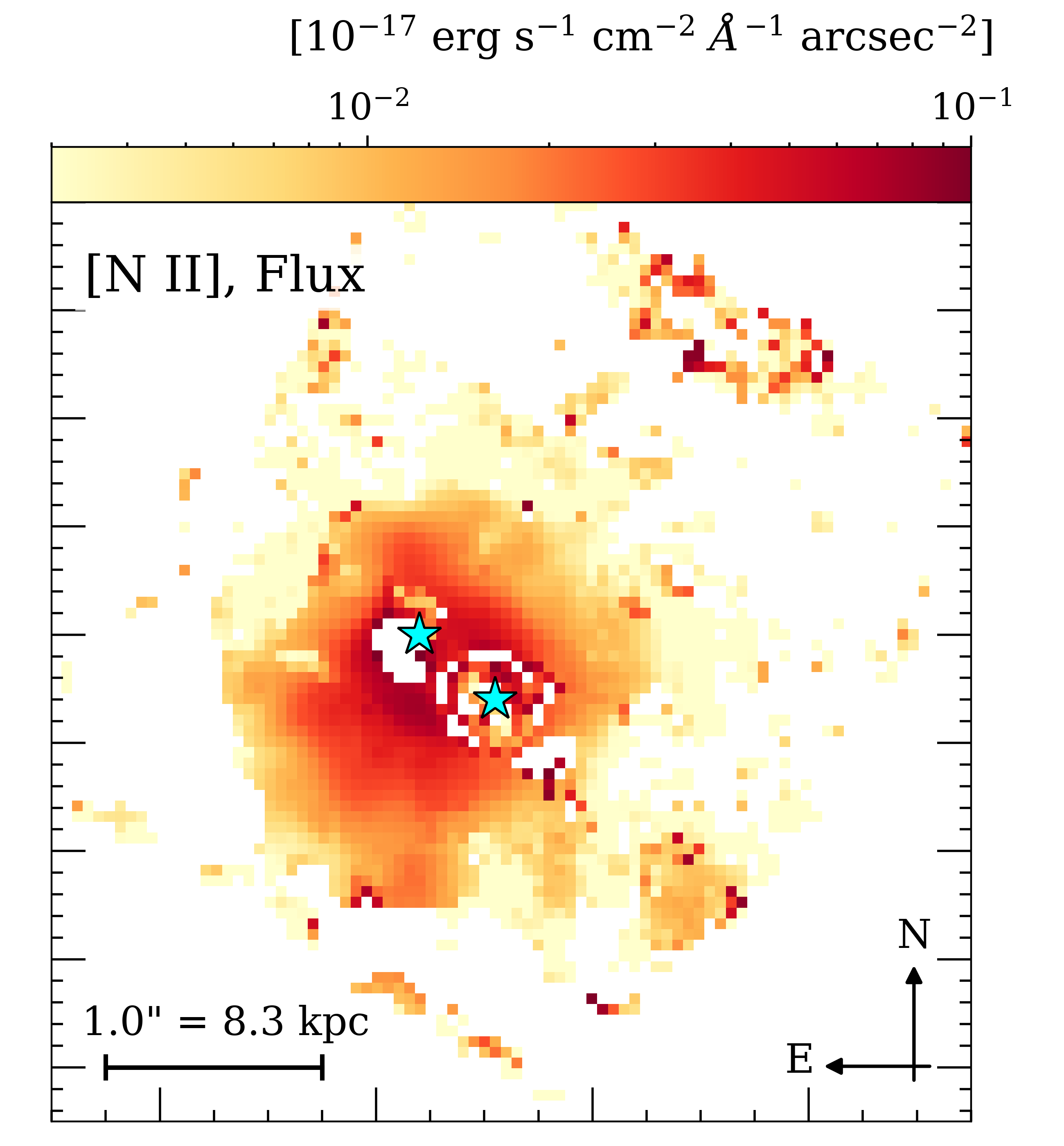}&
    \includegraphics[trim={0.3cm 0.3cm 0.4cm 0cm},clip,width=0.32\textwidth]{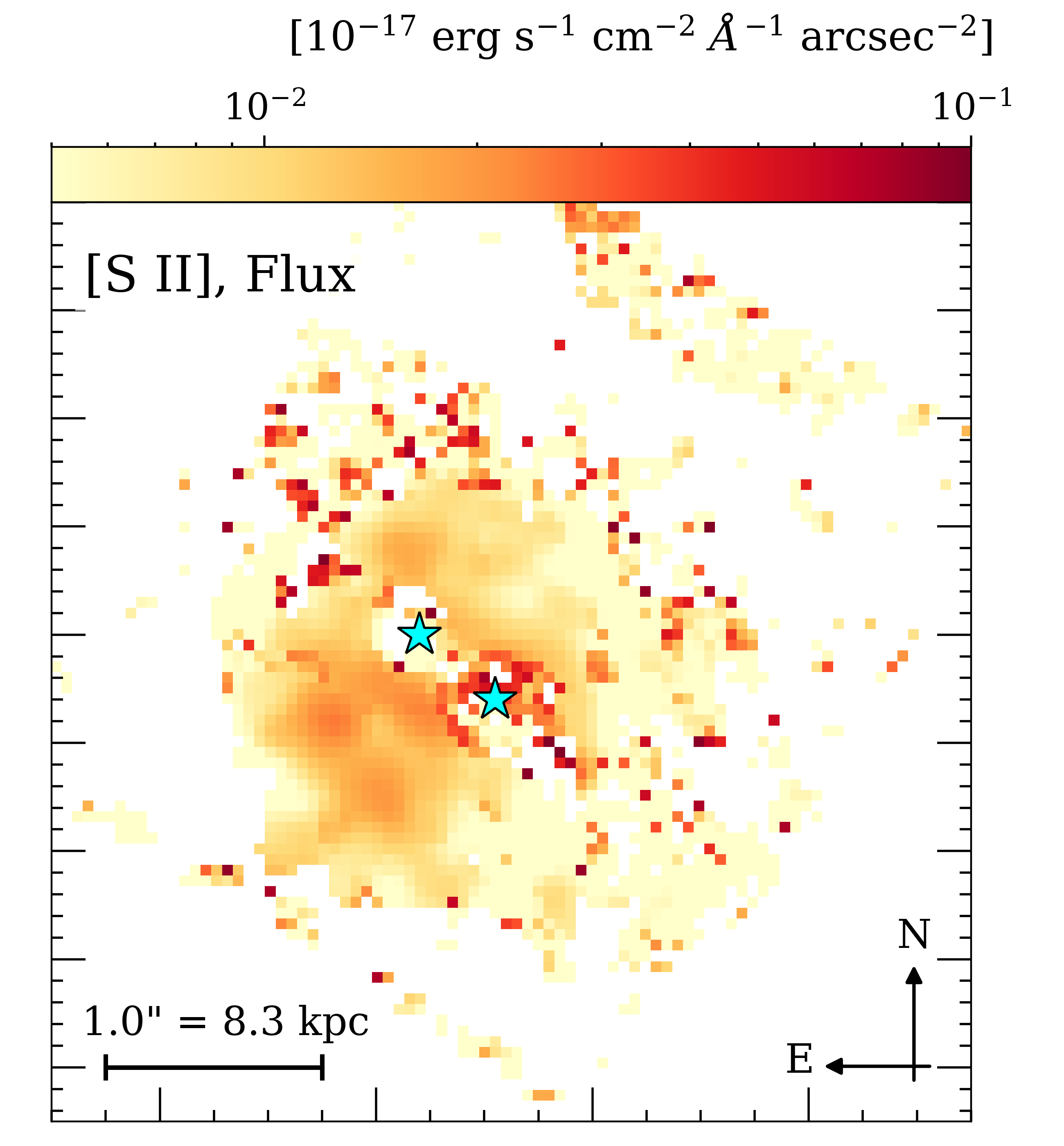}&
    \includegraphics[trim={0.3cm 0.3cm 0.4cm 0cm},clip,width=0.32\textwidth]{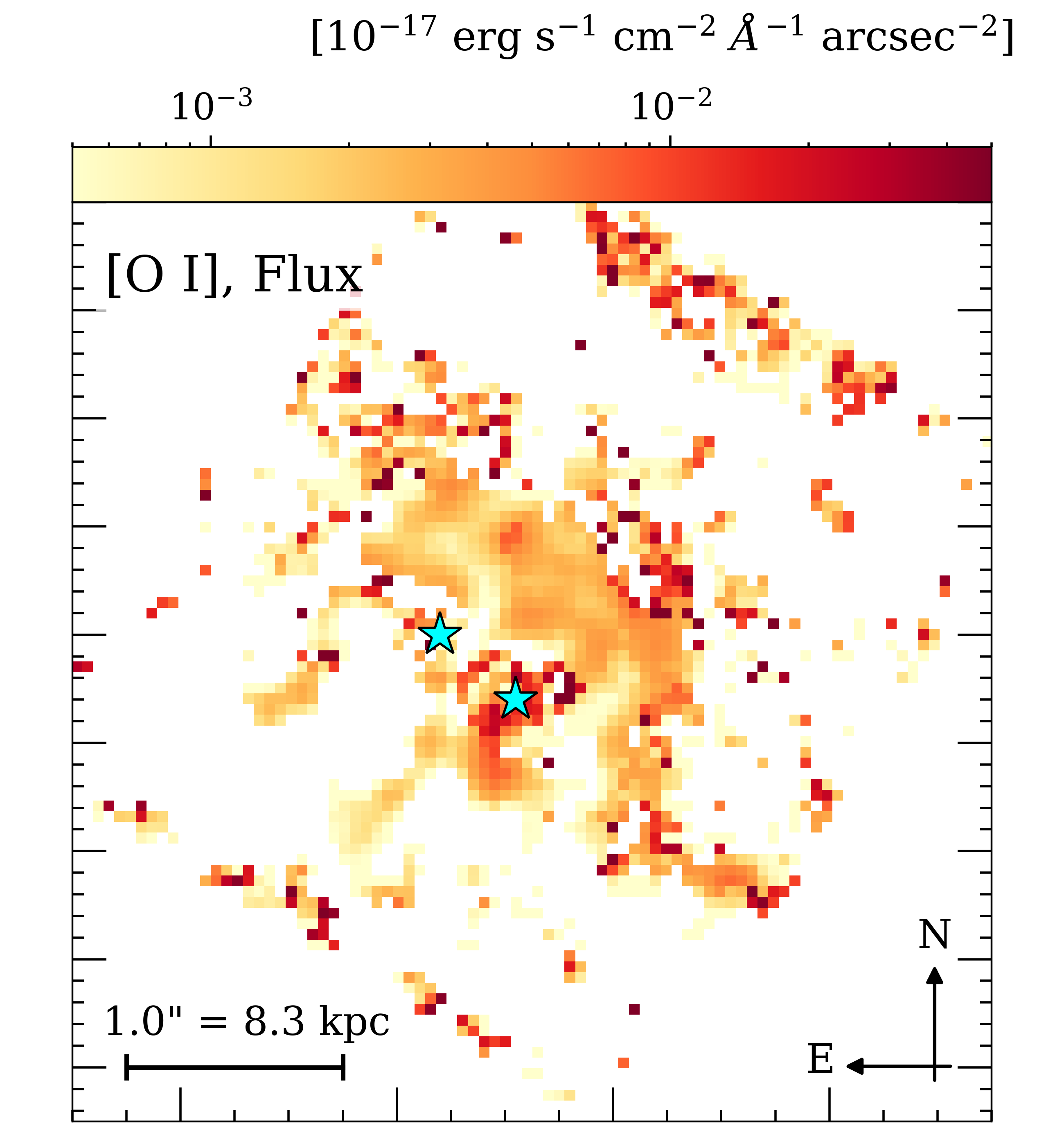}
	 \end{tabular}
	 \end{center}
	 \caption{Intensity maps of \nii\ (left), \sii\ (center), and \oi\ (right) produced with \qdfit. These lines are kinematically tied to \ha. We see that \nii\ is more centrally concentrated, whereas \sii\ and \oi\ have uneven distributions around the quasars compared to \ha\ in Figure \ref{fig:map_ha}.} 
	 \label{fig:maps_niisiioi} 
\end{figure*}

\begin{figure*}
	 \begin{center}
  \begin{tabular}{ccc}
    \includegraphics[trim={0.3cm 0.3cm 0.4cm 0cm},clip,width=0.32\textwidth]{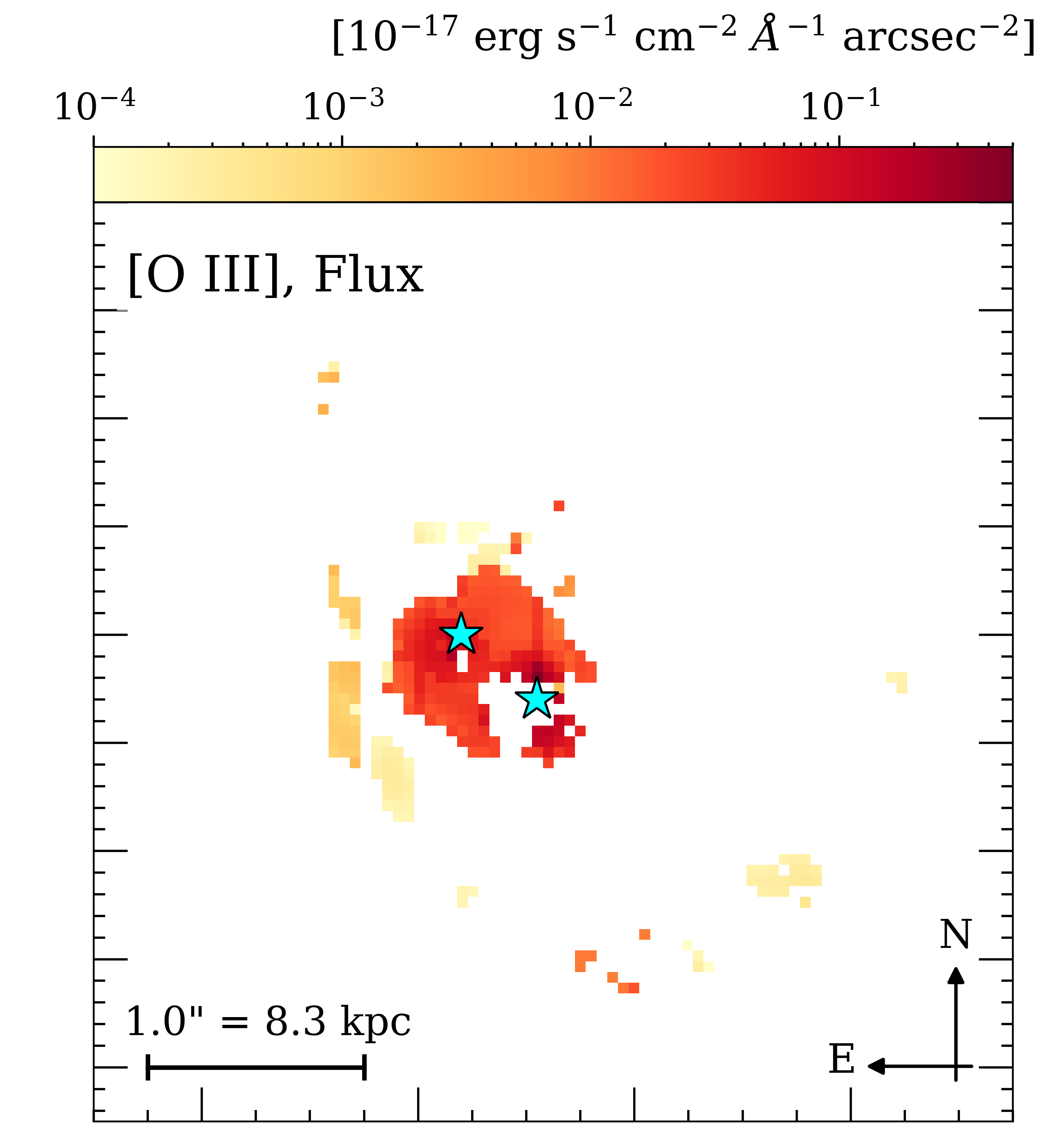}&
    \includegraphics[trim={0.3cm 0.3cm 0.4cm 0cm},clip,width=0.32\textwidth]{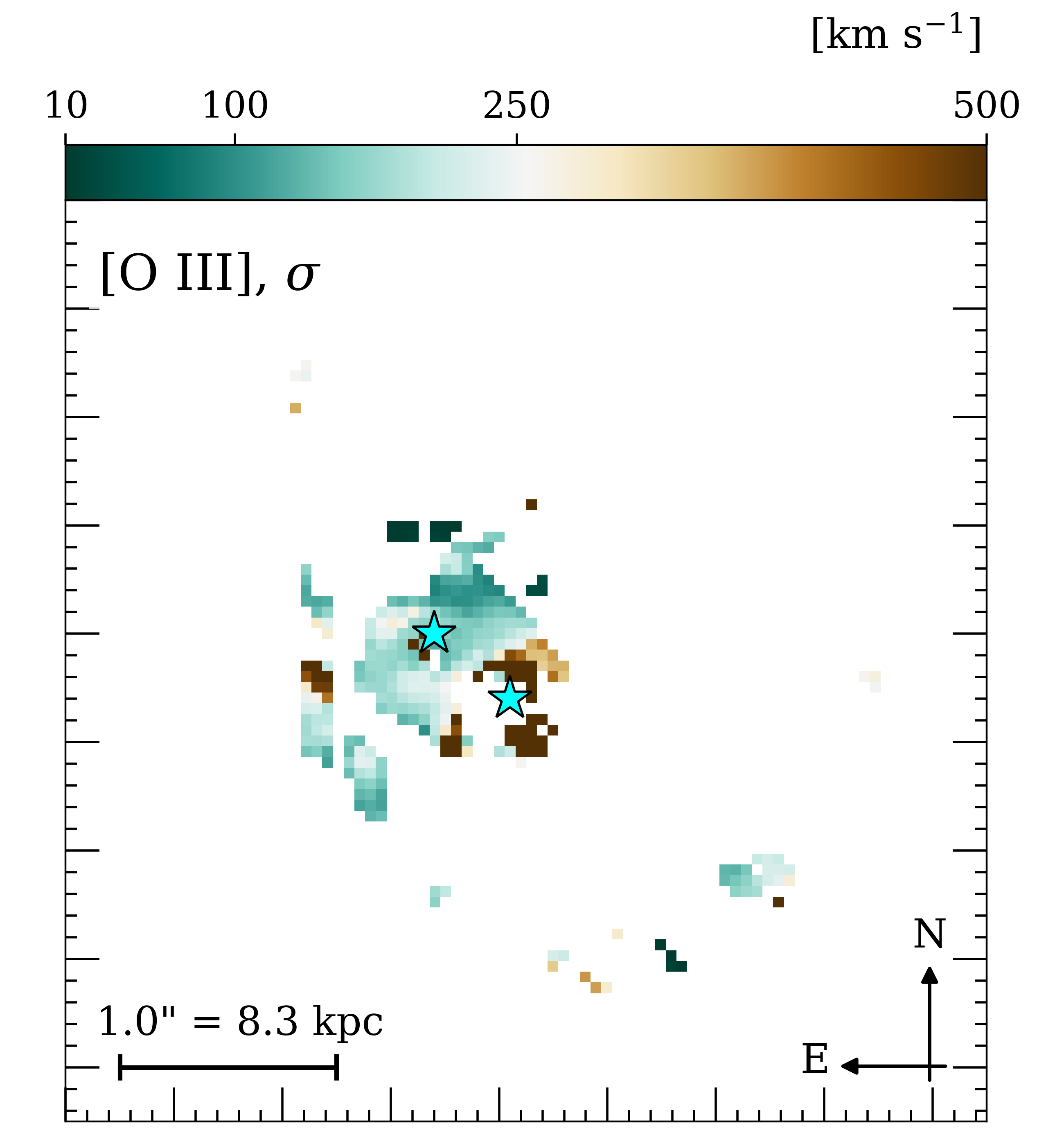}&
    \includegraphics[trim={0.3cm 0.3cm 0.4cm 0cm},clip,width=0.32\textwidth]{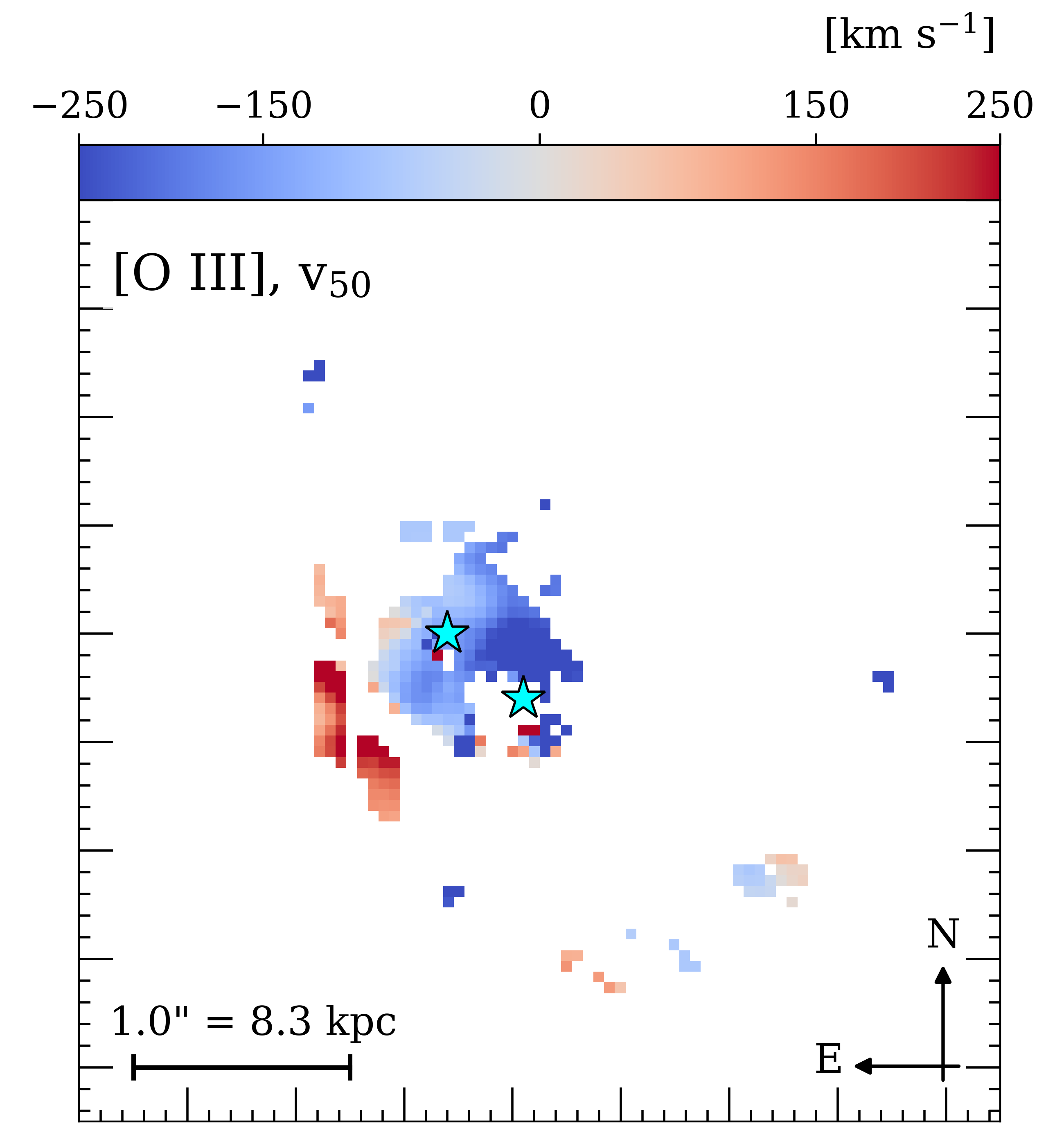}
	 \end{tabular}
	 \end{center}
	 \caption{Intensity (left), velocity dispersion (center), and velocity (right) maps of the quasar-subtracted, narrow-line \oiii\ produced with \qdfit. Compared to the \ha\ maps, we see that \oiii\ has a much smaller spatial extent.} 
	 \label{fig:map_hbo3} 
\end{figure*}

We run two separate fits for each grating setting. We fit the entire cube with a single component Gaussian components for each emission line (G140M to fit \hb\ and \oiii; and G235M to fit \ha, \nii, \sii, and \oi$\lambda6300$\AA). The \oiii\ and \nii\ line ratios are fixed. We assume that each Gaussian component is ``kinematically tied'' \citep{Zakamska2016a} across all emission lines for a given grating setting (e.g.~\hb\ and \oiii\ in G140M are tied together, but not with \ha\ in G235M). Successful detection of an emission line is set by two criteria: $>3\sigma$ peak intensity and the line widths that are greater than the instrumental width of the line-spread function. We repeat this for each quasar template.

Lastly, we combine the host galaxy fits corresponding to the two quasar templates. We iterate through each spaxel, compare the \qdfit\ line-fits for model-SW and model-NE, pick out the fits with the minimal $\chi^2$, and save them to the master datacube. 

\subsubsection{Host galaxy gas kinematics}
We recover diffuse emission of narrow lines that are best traced by the luminous \ha\ extending nearly $2.5\arcsec$ in diameter or $\sim20\ \textrm{kpc}$ measured across the center of the system. In Figure \ref{fig:map_ha}, we show the intensity, velocity dispersion, and radial velocity maps for \ha. 

For the first time, we detect the faint diffuse emission and measure its kinematics surrounding a close-separation dual quasar at $z\sim 2$. We see maximal radial velocity difference of $\Delta v\sim 1000 \textrm{ km s}^{-1}$ along the line perpendicular to the quasar alignment (northwest vs.~southeast regions) with a blue-shifted component to the northwest and a redshifted component to the southeast. The measured differences in the radial velocity of the diffuse gas emission are significantly larger than the velocity offset observed between the two quasars. Selected aperture spectra of \ha\ are shown in Figure \ref{fig:totalMAP}. The inferred $v_{50}$ velocity map suggests gas rotation with no obvious signs of kinematic disturbance often seen in major mergers. The quasars lie in the line of nodes of the galactic rotation field, whereas in a merger we expect each quasar to occupy either the red- or blue-shifted sides.

\begin{figure*}
	 \begin{center}
      \begin{tabular}{ccc}
    \includegraphics[trim={0.3cm 0.9cm 0.4cm 0.8cm},clip,width=0.32\textwidth]{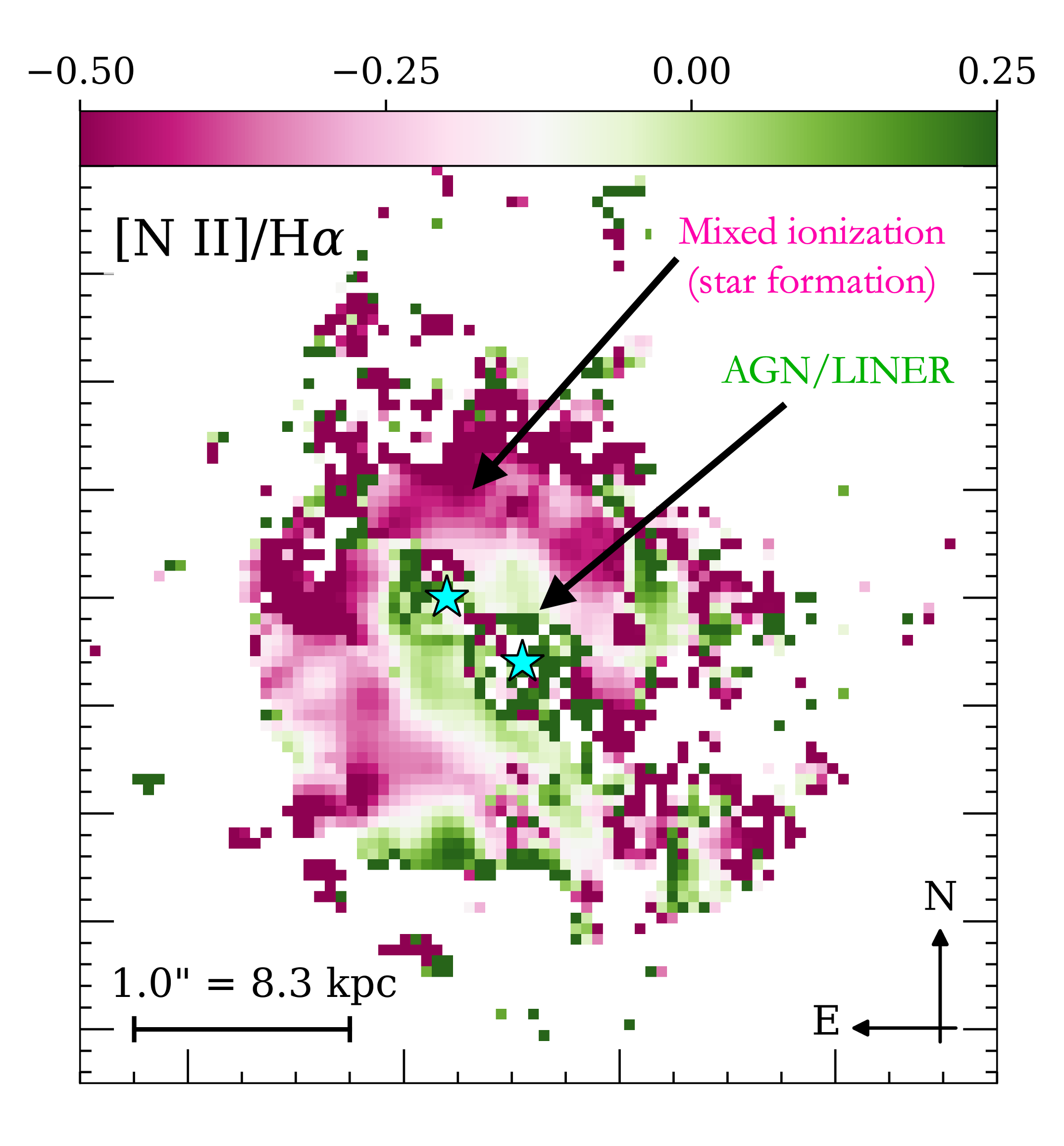}&
    \includegraphics[trim={0.3cm 0.9cm 0.4cm 0.8cm},clip,width=0.32\textwidth]{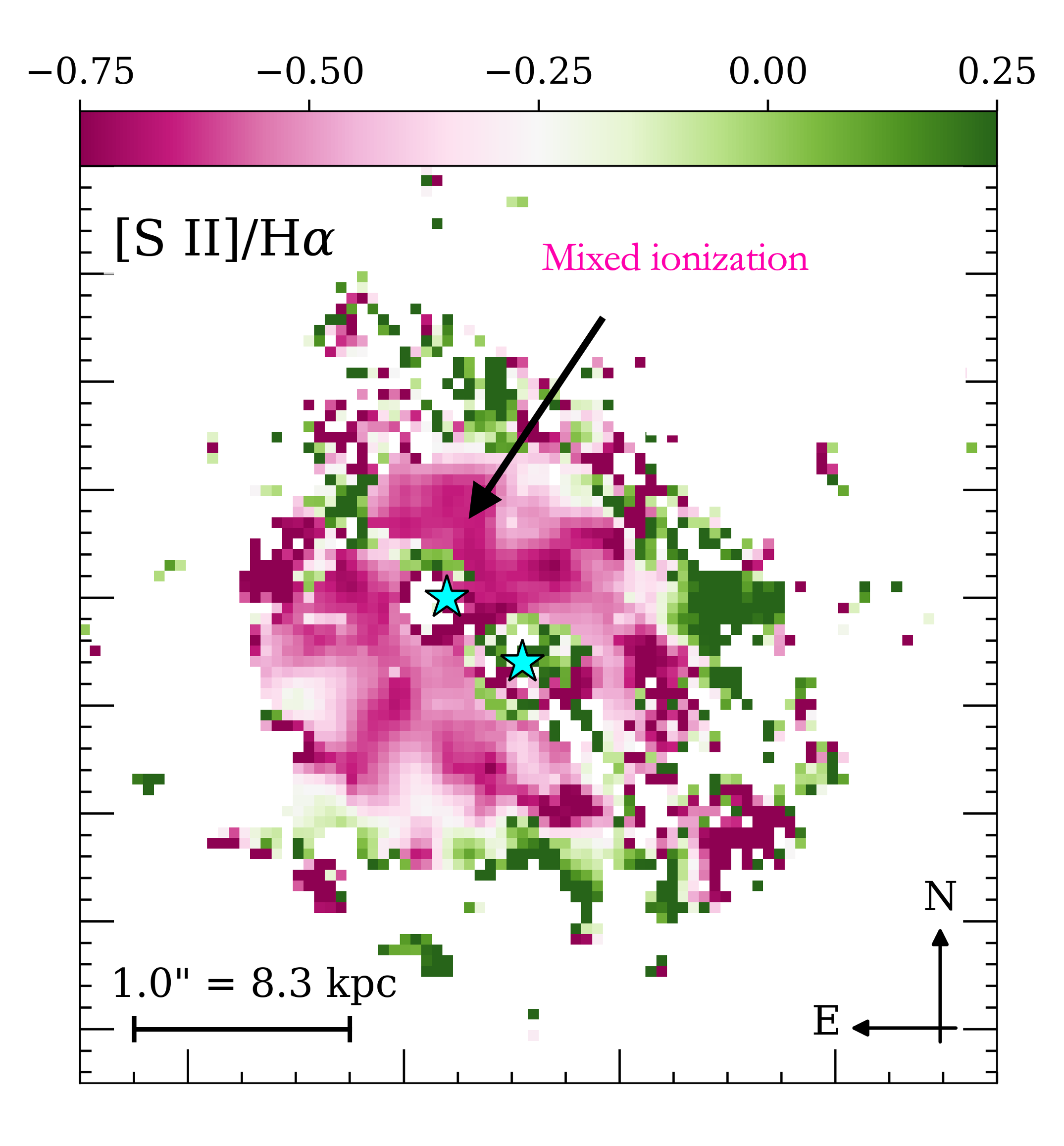}\\
    \includegraphics[trim={0.3cm 0.9cm 0.4cm 0.8cm},clip,width=0.32\textwidth]{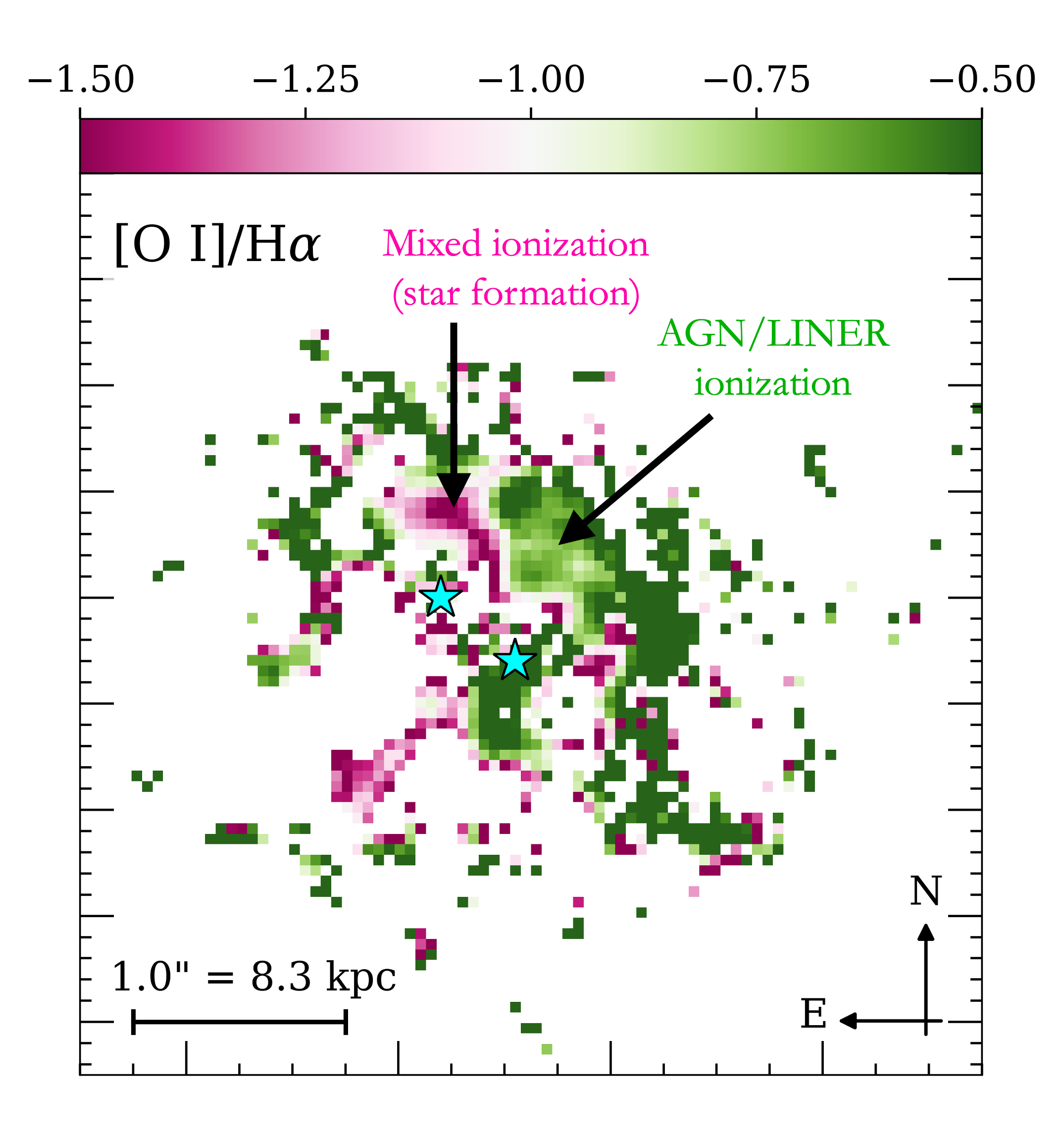}&
    \includegraphics[trim={0.3cm 0.9cm 0.4cm 0.8cm},clip,width=0.32\textwidth]{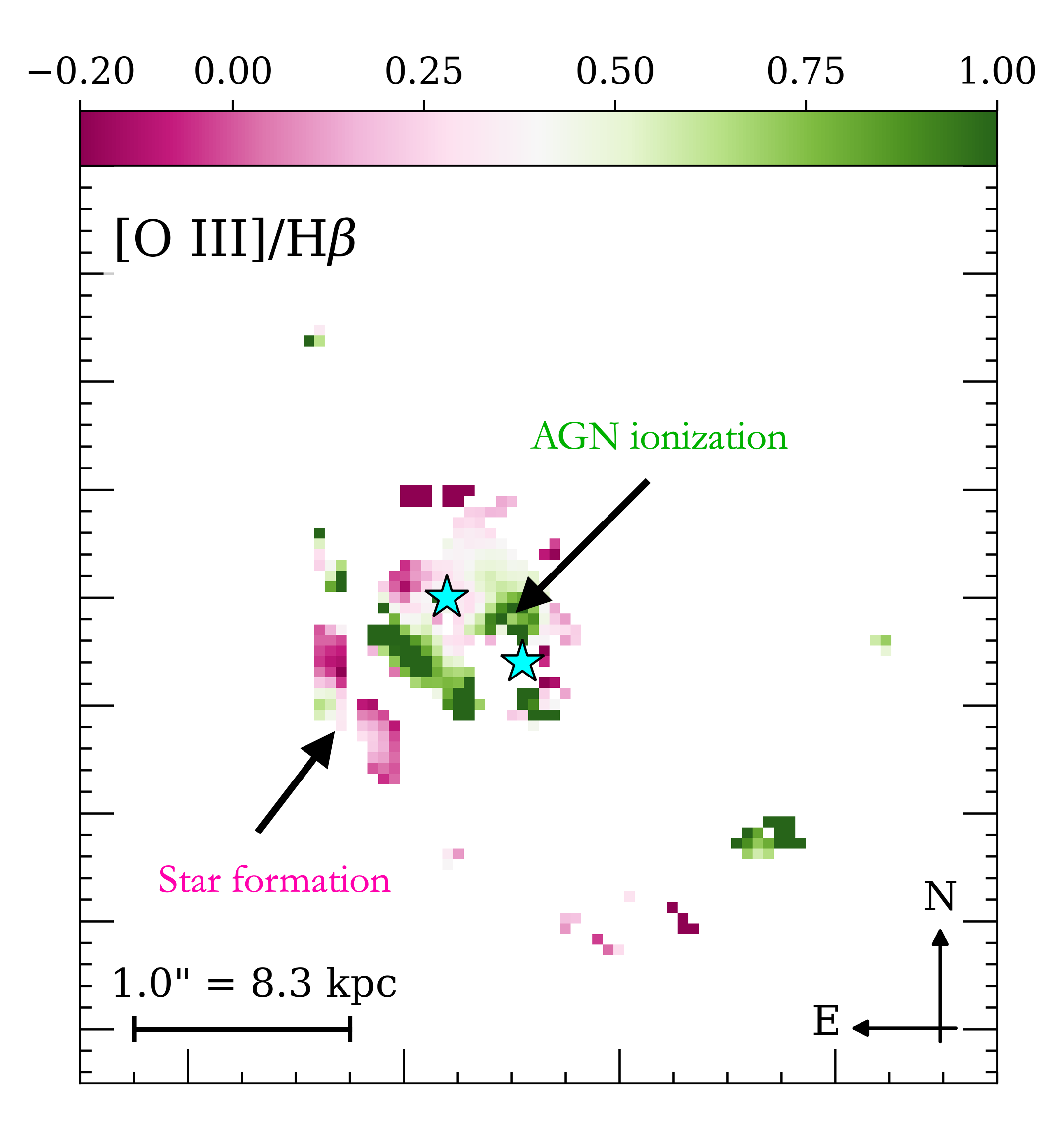}&
    \includegraphics[trim={0.3cm 0.9cm 0.4cm 0.3cm},clip,width=0.29\textwidth]{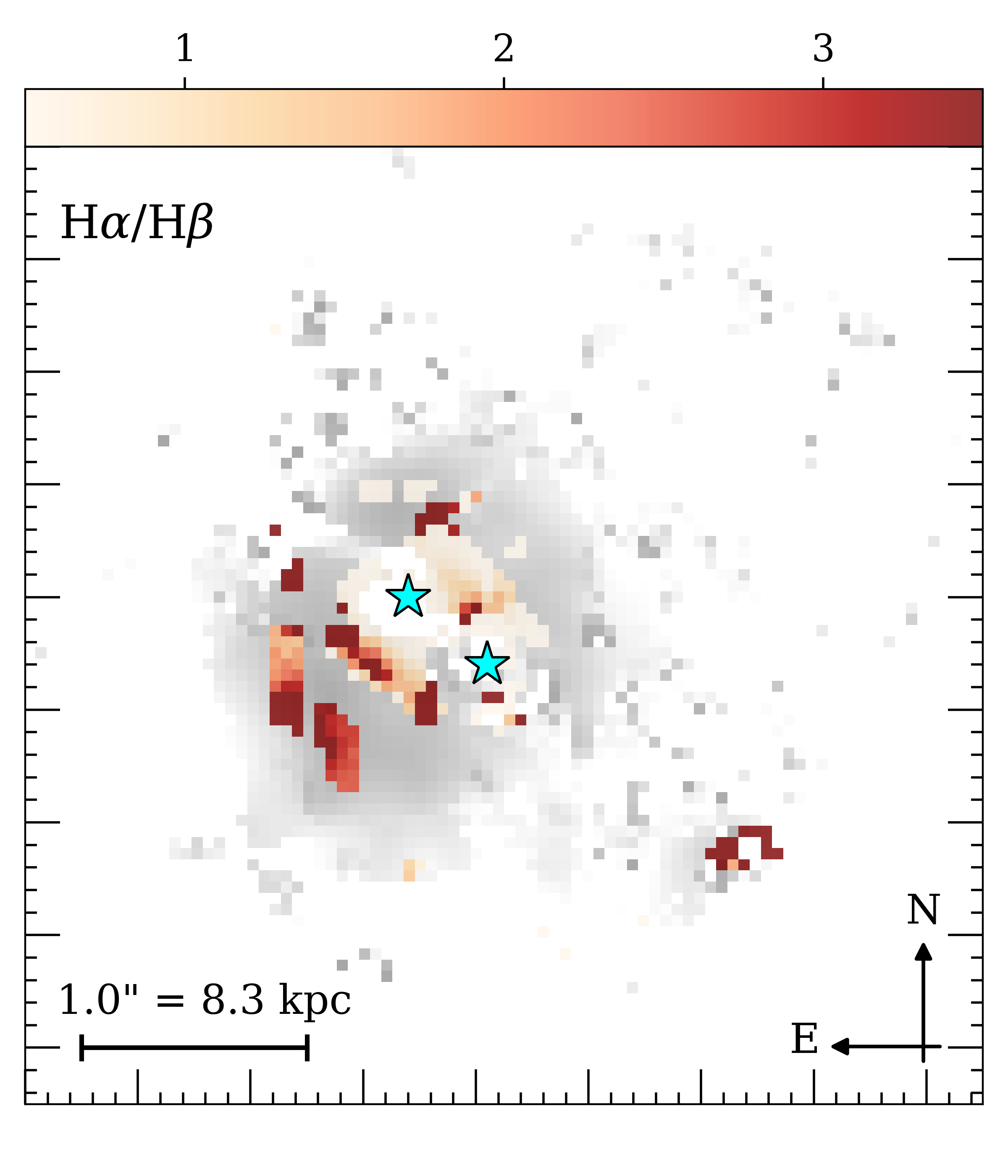}
	 \end{tabular}
	 \end{center}
	 \caption{We show the $\log_{10}$\niiha\ (top left), $\log_{10}$\siiha\ (top right), $\log_{10}$\oiha\ (bottom left), $\log_{10}$\oiiihb\ (bottom center), and \HaHb\ (bottom right) line ratio plots. The green zones indicate regions with increased AGN- and/or LINER-like ionization, whereas the magenta zones show mixed ionization from star formation and the quasars. The \oiiihb\ map suggests clumps of star formation concentrated around \targNE\ and to the east. The enhanced quasar ionization around \targSW\ is also supported in \niiha. The \siiha\ and \oiha\ ratios suggest a more mixed ionization in the center. We show the corresponding BPT and VO87 diagnostics in Figure \ref{fig:iondiag}. The \HaHb\ Balmer decrement map (in red/orange gradient), which shows little dust attenuation, is overplotted over the extended \ha\ map shown in gray.} 
	 \label{fig:lineratio} 
\end{figure*}

In addition to the narrow \ha\ emission, we also detect faint extended narrow line \hb, \oiii, \oi, \nii, and \sii, although they are not as luminous or extended as \ha. We show the intensity maps of the narrow \nii, \sii\, and \oi\ in Figure \ref{fig:maps_niisiioi}. We also detect a \ha\ and \nii\ bright T1 companion to the southwest \citep{ChenYC2023a} with a mean velocity shift of $-120\ \textrm{km s}^{-1}$. 

In Figure \ref{fig:map_hbo3} we show the intensity and kinematic maps of \oiii. Both \nii\ and \sii\ are co-spatial with the extended \ha\ that surround the two quasars. The \oi\ emission is mostly concentrated in the northwest. In contrast, \hb\ and \oiii\ emission is more compact and clumpy, mostly centered around \targNE, as seen in Figure \ref{fig:map_hbo3}. Interestingly, \oiii\ is fainter around \targSW, despite it being the more luminous of the two quasars. The velocity structure of \oiii\ is also slightly different from that of \ha\ with a greater number of blueshifted clumps dominating the regions between \targNE\ and \targSW. Although there are regions with large \oiii\ velocity offsets reaching $\Delta v\sim400\textrm{ km s}^{-1}$, the associated velocity dispersion is small, so it is unlikely that \target\ has any large-scale outflows such as those found in recent JWST observations of luminous quasars at cosmic noon \citep[e.g.,][]{Wylezalek2022, Vayner2023, Veilleux2023}. 

\subsubsection{Host-galaxy Ionization}

We use optical line diagnostics \citep{Baldwin1981, Veilleux1987} to investigate the ionization mechanisms of the observed line emission, hereafter BPT \citep{Baldwin1981} and VO87 \citep{Veilleux1987}. We take the emission line maps and measure the \niiha, \siiha, \oiha, and \oiiihb\ line ratios, shown in Figure \ref{fig:lineratio}. The detected distribution of \oiiihb\ is more compact, due to their inherent faintness compared to the \ha\ emission. We also map the \HaHb\ line ratio to estimate the Balmer decrement. 

\begin{figure*}
	 \begin{center}
      \includegraphics[trim={0cm 0.5cm 0cm 0cm},clip,width=0.99\textwidth]{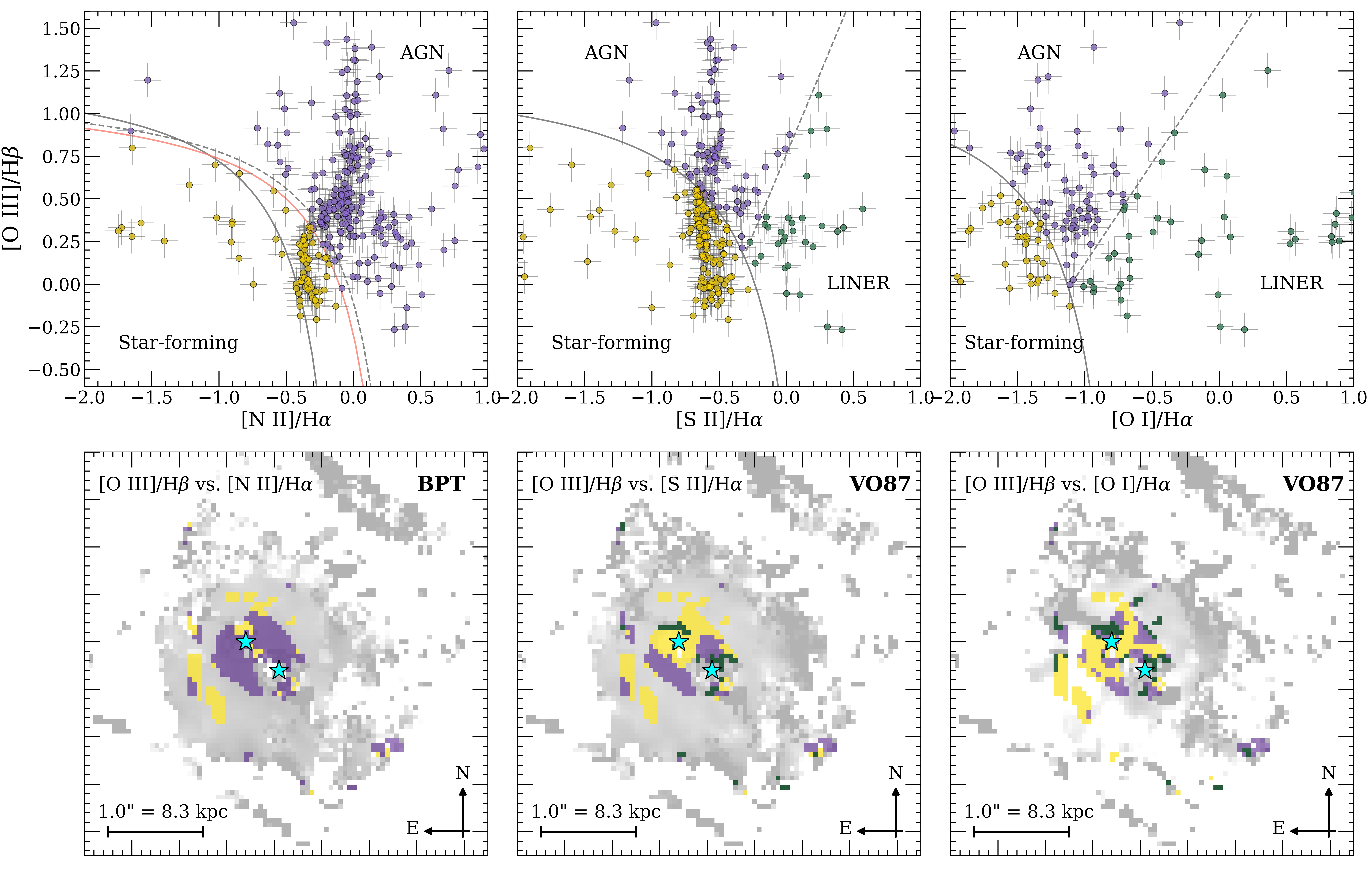}
	 \end{center}
	 \caption{We show the spatially-resolved \oiiihb\ vs.~\niiha\ BPT (top left), \oiiihb\ vs.~\siiha\ VO87 (top center), and \oiha\ vs.~\siiha\ VO87 (top right) ionization diagrams. We show the line that separates star formation and AGN/quasar photoionization. The BPT diagram (top left) shows the solid gray line ($z=0$ model; \citealt{Kauffmann2003}), dashed gray line (maximal $z=3$ model; \citealt{Kewley2001}), and solid red line ($z=2.17$ model; \citealt{Kewley2013b}). The VO87 diagrams (top center and top right) shows the solid gray line separating star formation and AGN photoionization, and a dashed line separating quasar and LINER-like shock ionization. The points on the BPT and VO87 diagrams are color-coded to match the corresponding spatial positions shown in the bottom row: (bottom left) BPT map and (bottom center and right) VO87 map. The gray regions mark the \niiha, \siiha, and \oiha\ fluxes from Figure \ref{fig:lineratio} overlaid with quasar photoionized (purple), star-forming (yellow), and shock ionized (green) regions.} 
	 \label{fig:iondiag} 
\end{figure*}

We show the BPT (\niiha\ vs.~\oiiihb) and VO87 (\siiha\ vs.~\oiiihb\ and \oiha\ vs.~\oiiihb) measurements along with the theoretical lines delineating photoionization by star formation, quasar, and shocks in Figure \ref{fig:iondiag}.  Studies \citep[e.g.][]{Sanders2015,Strom2017, Strom2018, Sanders2023} show that at higher redshifts, the effects of harder ionization spectra, lower gas and stellar metallicities, and denser interstellar medium push the separation between the star formation and active galactic nuclei (AGN) or quasar ionization above the classic $z=0$ line \citep{Kauffmann2003} in the BPT diagram. We adopt the \citet{Kewley2013b} formalism to calculate star formation in \target. Using the \cite{Kewley2013b} criteria at $z=2.17$, we map the associated star for ionization regions in Figure \ref{fig:iondiag}. We also plot the BPT and VO87 diagrams and map the associated ionization regions.

We can see that the gas in \target\ is a mix of quasar photoionization (in purple) and star formation (in yellow) with some shock ionization (in green) hinted from elevated \oi\ emission close to the quasars. The extended regions are mostly dominated by star formation. Although ionization at large distances from the quasar is likely due to star formation, it is difficult to determine the exact ionization sources due to the faint nature of \hb\ and \oiii\ at these distances. Both the BPT and VO87 diagnostics indicate that regions around \targNE\ and the clumps to the southeast are associated with star formation, while regions around \targSW\ is associated with quasar ionization. The southeastern star-forming clumps are associated with high \HaHb\ suggesting dusty star formation.

Considering the high luminosities of the two quasars ($L_{bol}\sim10^{46}\textrm{ erg s}^{-1}$; Table \ref{tab:fitproperties}), it is not surprising that the inner regions to be dominated by quasar photoionization. It is more intriguing to find streams of star-forming regions centered on \targNE. There are some indications of star-forming streams connecting the two quasars towards the northwest (blueshifted \ha). The star-forming streams are best traced by the \siiha\ and \oiha\ line ratio diagnostics, shown in Figure \ref{fig:iondiag}. The extended \nii\ and \sii\ emission also show low \niiha\ and \siiha, although the \hb\ and \oiii\ are not detected. Since there is no evidence for widespread quasar-driven outflows, the likely explanation for the extended \ha, \nii, and \sii\ is ionization due to low levels of star formation; the \ha\ surface brightness in these regions are nearly $\times10$ fainter than the central regions. The total \ha\ flux associated with these star-forming regions is $6.0\times10^{-15} \textrm{ erg s}^{-1}\textrm{ cm}^{-2}$, which translates to $\textrm{SFR}\sim1,700\ M_{\odot}\textrm{ yr}^{-1}$ based on the empirically derived conversion between \ha\ luminosity and SFR \citep{Kennicutt1998}. If we account for additional contribution from quasar photoionization in the central regions, this SFR estimate may be an upper limit, but it is clear that the SFR is very high. Assuming a stellar mass of $10^{11.78}\ M_{\odot}$ \citep{ChenYC2023a}, we obtain a specific SFR of $2.7\ \textrm{ Gyr}^{-1}$. The results are qualitatively in agreement with the recent finding by \citet{ChenYC2024} based on the \jwst/MIR data that the host galaxy of \target\ is undergoing very active star formation. 

In addition to the star-forming streams, the elevated \siiha\ and \oiha\ line ratios indicate pockets of LINER-like ionization (green in Figure \ref{fig:iondiag}). These shock regions are concentrated in the central regions along the star-forming streams and the quasar ionization, which may suggest shock heating. These regions also show elevated velocity dispersion reaching $\sigma\sim150\textrm{ km s}^{-1}$ but show modest velocity offsets of $\sim100\textrm{ km s}^{-1}$. The likely culprits of the observed shock signatures are ionization from star formation (or stellar feedback) in the dense interstellar medium or tidal shocks due to a merger \citep{Rich2015}. Quasar-driven outflows are unlikely based on the \oiii\  kinematics. 

The Balmer decrement of the broad \ha\ and \hb\ lines indicate little reddening or obscuration close to the quasar. However, the narrow line ratio in select clumps to the southeast is higher. Incidentally, BPT diagnostics indicate star formation dominates in these regions. Assuming the $R_V=3.1$ dust extinction model \citep{Cardelli1989}, we obtain a range of extinction up to $A_V\approx1.75\textrm{ mag}$. In Section \ref{sec:specAnaly} we noted that the \targNE\ quasar has a slightly redder continuum than that of \targSW, but this reddening is not evident in the extended line emission. This may suggest that the nuclear continuum reddening is a local effect rather than at galactic scales. 

\section{Dual quasar or lensed quasar?} \label{sec:disc:duallens}
Distinguishing between lensed quasars and dual quasars is particularly challenging, especially for small-separation pairs at higher redshifts. Although \cite{ChenYC2023a} previously ruled out gravitational lensing in \target, the new JWST observations of the two quasars revealed surprisingly similar spectra, prompting a re-evaluation of the lensing hypothesis. This section compares the evidence supporting the dual quasar hypothesis vs.~the gravitational lensing hypothesis.

\subsection{Evidence for dual quasar:} 
As outlined in Section \ref{sec:prevOBS}, \cite{ChenYC2023a} initially ruled out lensing based on several factors, including the non-detection of a foreground lens source, the tentative detection of tidal tails from the host galaxy, chromatic differences in the X-ray and radio emission, and subtle differences in the quasars' optical spectra. Strong lensing modeling of the optical images disfavored a lensing scenario. Figure \ref{fig:multiwave_qso_compare} presents the rest-frame UV to optical spectral energy distribution from these observations (SDSS, HST, Gemini, Chandra, VLB-A) along our new JWST/NIRSpec data. 

Detailed analyses of the quasar spectra in Section \ref{sec:specAnaly} revealed subtle yet clear differences in the continuum shape, emission line profiles (e.g.~\hb, \oiii, \ha, \hei), and the \feii\ emission complex. \pyqsofit\ modeling of the emission lines identified clear differences in the line shapes: \targNE\ exhibits a weaker broad-line component and stronger narrow-line emission compared to those of \targSW, resulting in ``pointier'' spectral profiles as shown in Figure \ref{fig:pyqsofit}. There is also a divergence in the quasar continuum redward of \ha\ as shown in the flux ratio in Figure \ref{fig:multiwave_qso_compare}, which is consistent with the differences in the rest-frame UV continuum slope measured by \cite{ChenYC2022}. The difference in slope does not correspond to the variations of \HaHb\ in Figure \ref{fig:iondiag}, which may be an argument against differential reddening by a foreground lens galaxy. We detect a velocity offset of $\Delta v \sim 200 \textrm{ km s}^{-1}$ between the two quasars. 

Perhaps the more compelling evidence from JWST is the detection of the faint, extended host galaxy traced by multiple narrow emission lines (e.g.~\oiii, \ha, \hb) at the same redshift as the two quasars. The \ha-emitting region spans nearly $2\arcsec$ or $16\ \textrm{kpc}$ in diameter, with kinematic analysis suggesting a large rotating disk. Furthermore, the morphology and distribution of the extended ionized gas emission do not resemble strong lensing features such as arcs or symmetrical structures. Moreover, the spatial distribution of the different line-emitting gas is different. For instance, \ha\ is distributed nearly uniformly across the field, while the \oiii\ emission is concentrated around \targNE\ and the \oi\ emission is concentrated to the northwest of the quasar pair (see Figures \ref{fig:map_ha}, \ref{fig:maps_niisiioi}, and \ref{fig:map_hbo3}). The flux ratios of the extended emission lines are inconsistent with that of the two quasars. 

Further support against lensing includes the absence of quasar image flipping or mirrored kinematics expected from lensing parities. Instead, we see one normal-looking rotating disk with a symmetric red-/blueshifted kinematic map. The spectral differences in the emission lines also rule out scenarios involving scattered light producing duplicate quasar images. Furthermore, there is no evidence of absorption line features, such as \mgii\ and \feii\ doublet, associated with a foreground lens galaxy, such as those seen in \cite{Gross2023}. Lastly, we have yet to find any indication of a foreground lens -- either through HST imaging or JWST spectroscopy. 

\subsection{Evidence for lensing:} One significant and non-negligible evidence in favor of the lensing hypothesis is the striking similarity between the two quasars, despite the detection of suble emission line differences. Although we measure velocity offsets between the two quasars' spectra of up to $\Delta v\sim200\ \textrm{km s}^{-1}$, the uncertainties render these measurements inconclusive. Consequently, lensing remains a plausible and natural explanation for the observed quasar similarity, especially given the inherent challenges in robustly interpreting the velocity offset. 

Various other observables in favor of the dual quasar hypothesis may be explained by lensing as well. The observed continuum reddening may be attributed to differential reddening caused by a foreground lens galaxy \citep[e.g.,][]{Sluse2012, Agnello2018}. Subtle differences in the emission line profiles may also be explained by lensing-induced time delays. Notably, although no lensing galaxy has been detected, this non-detection does not definitively rule out the lensing hypothesis. For example, \cite{Hawkins1997a}, among others, reported the discovery of a lensed quasar pair with a so-called ``dark'' lens galaxy, although the lensing galaxy was later identified with improved photometry and imaging analysis \citep{Hawkins2021}. Thus, more exotic lensing scenarios, such as an extremely faint, massive lens (e.g.~ultra-diffuse galaxy) or even a dark lens, cannot be entirely excluded. The lack of kinematic distortions in the gas, as revealed by JWST IFU observations at face value, is inconsistent with expectations for a major merger, as further detailed in Section \ref{sec:disc:merger}. 

Ultimately, the evidence for both dual quasars and lensing remains circumstantial, despite the multi-wavelength observations \citep[e.g.][]{Shen2021, ChenYC2023a} and spatially-resolved spectroscopy with JWST (this work and \citealt{ChenYC2024}). This underscores the inherent challenges in not only identifying but also conclusively confirming the nature of dual quasar candidates. If \target\ is indeed a physical pair, then their quasar spectra may reflect interesting accretion physics, perhaps unique to dual quasars at the observed separation and merger stage. We explore these possibilities further in the following sections, assuming the dual quasar interpretation. 

\section{Interpreting the dual quasar case}\label{sec:disc:dualBH}
\subsection{Merger or massive disk galaxy?}\label{sec:disc:merger}

The detection of extended gas emission associated with the host galaxy environment provides compelling evidence supporting the dual quasar hypothesis, particularly given the weak evidence of the nuclear velocity offset. With JWST, we are, for the first time, able to detect the host galaxy gas in \target\ traced by the extended narrow emission lines (e.g.~\ha, \hb, \oiii). Perhaps the most puzzling observation of \target\ is two distinct quasars embedded in what appears to be a large rotating gas disk, as illustrated by the \ha\ $v_{50}$ map in Figure \ref{fig:map_ha}. To interpret this observation, we propose two possible scenarios: either \target\ represents an ongoing major galaxy merger with two active quasars, or it is a large, single-disk galaxy hosting dual quasars. 

\textbf{Galaxy merger?} The most natural scenario to form \target\ is through a major merger of gas-rich galaxies. We have two main evidence that supports the merger scenario: the detection of extended sources (e.g.~tidal tail) and the formation of a gas disk. 

Previously, \cite{ChenYC2023a} used \hst/F160W imaging to model \target\  with two quasar PSFs and two Sersic host galaxies; this analysis suggested an indirect detection of extended tidal tails, indicative of a merger. Some of these features (T3 and T4 in \citealt{ChenYC2023a}), located near the two quasars, are confirmed in the NIRSpec detection of \hb\ and \oiii, as shown in Figure \ref{fig:map_hbo3}. Likewise, T1 source to the southwest is confirmed, as shown in Figure \ref{fig:totalMAP} with a blueshift of $-120\ \textrm{km s}^{-1}$ that may either be a companion or a tidal tail feature. Based on the line ratios, T1 is \ha, \nii, and \sii\ bright, yet \hb+\oiii\ faint, so it is likely that T1 is dominated by star formation. 

Simulations also predict the formation of gas disks following a major gas-rich merger \citep{Barnes2002}, supporting the merger scenario. Interestingly, both observations and simulations suggest that the timescale for the presence of tidal tails and the formation timescale for gas disk formation overlap \citep[e.g.][]{Lotz2008, Dadiani2024}. Thus, the literature supports our interpretation of \target\ having both extended tidal tails and gas disk. 

Can the system ``hide'' a merger? Although kinematic maps can be used to distinguish between mergers and disks to some extent \citep{Wisnioski2015}, there are both observational and theoretical works highlight challenges to the method \citep{Simons2019, Nevin2021, Ciraulo2021}. Typically, a rotating disk galaxy exhibits a symmetric red-/blue-shifted $v_{50}$ velocity map, whereas mergers display more asymmetric kinematics \citep{Wisnioski2015}. However, \cite{Simons2019} demonstrated that the merger-disk classification for $z\sim2$ galaxies based on kinematic maps can be ambiguous, as mergers may masquerade as disk galaxies, especially in the absence of sufficient signal-to-noise and spatial resolution. Similarly, \cite{Nevin2021} showed that early-stage galaxy mergers can retain disk-like stellar kinematics, complicating their identification. One way to break this degeneracy may be to compare the morphological and kinematic (mis)-alignments between the stellar and ionized gas components \citep{BarreraBallesteros2015} in the future analysis of the stellar continuum data. Disentangling a ``hidden'' merger even at $z=0.09$ required multi-wavelength analysis, including observations of the stellar component \citep{Ciraulo2021}. We explore potential low redshift analogs to \target\ in Section \ref{sec:disc:lowz}.

\begin{figure*}
	 \begin{center}
	 \begin{tabular}{lc}
    \includegraphics[trim={0.4cm 0.3cm 0.4cm 0.3cm},clip, width=0.53\textwidth]{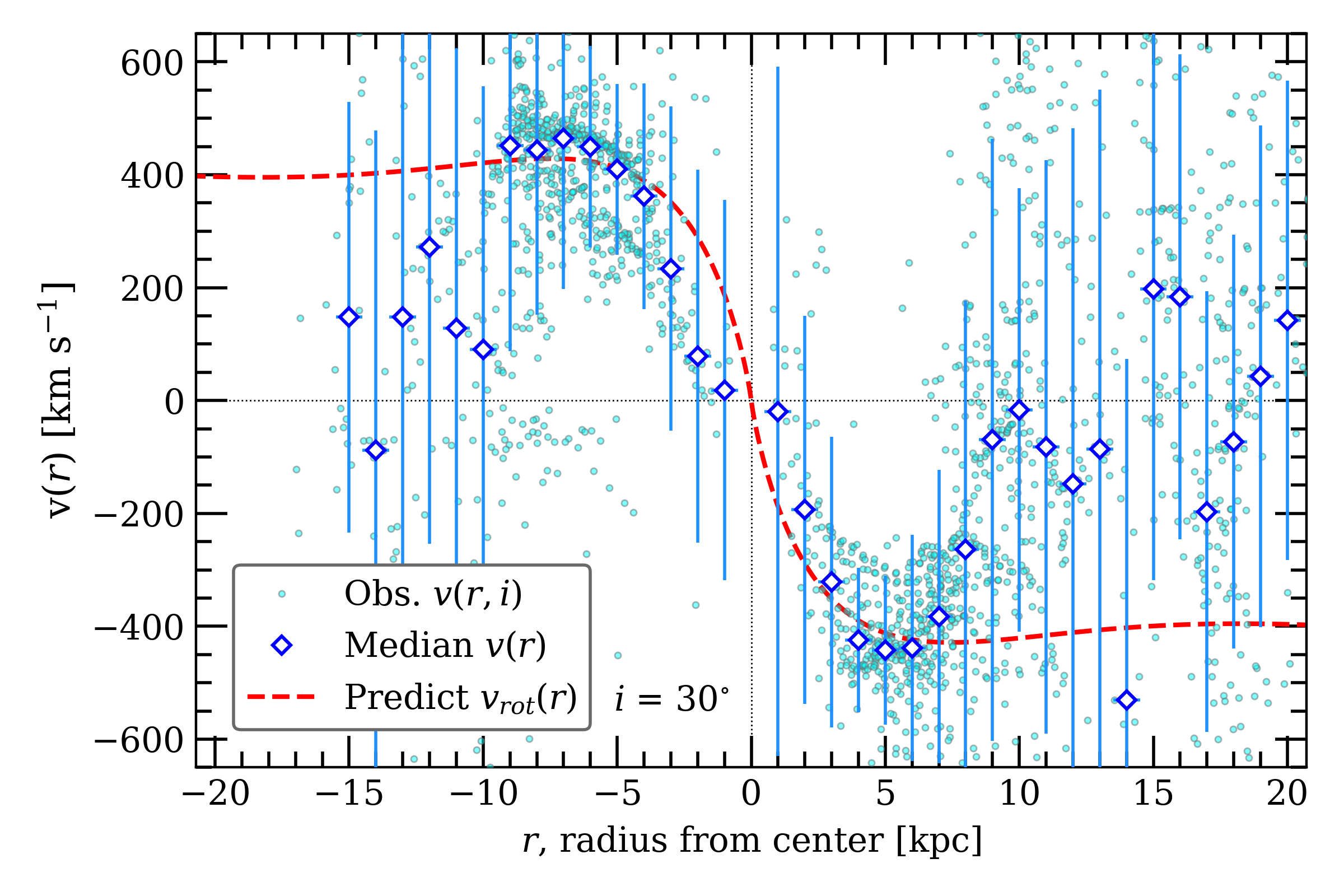} &
    \includegraphics[trim={0.6cm 0.3cm 0.5cm 0cm},clip, width=0.32\textwidth]{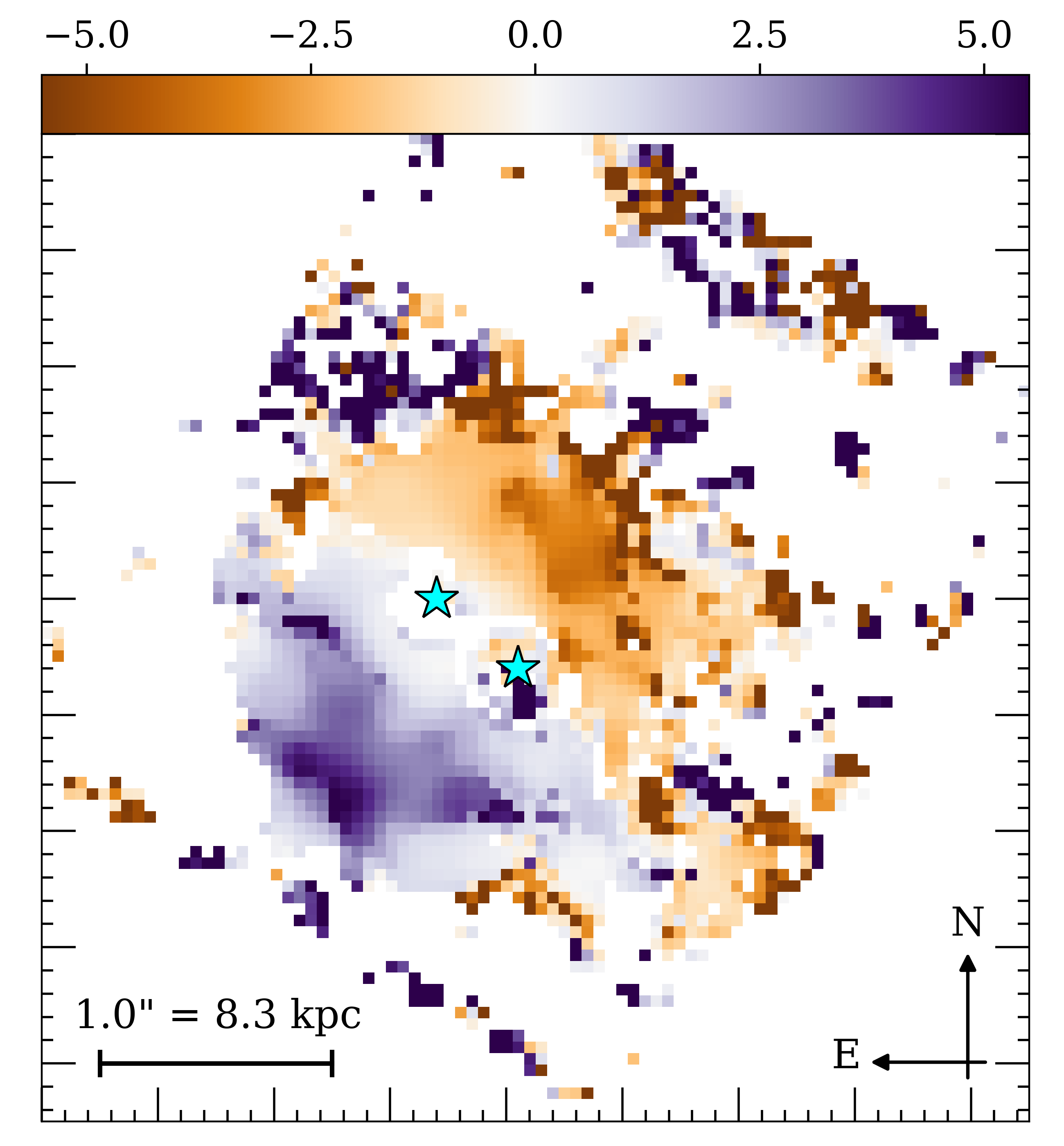} 
	 \end{tabular}
	 \end{center}
	 \caption{(Left) The deprojected $v(r)$ values, assuming an inclination angle of $i=30^{\circ}$,  based on the \ha\ kinematics are shown in small cyan circles. The median $v(r)$ values in each 0.5 kpc step are shown in blue diamonds with the error bars indicating the $1\sigma$ scatter. The \ha\ data suggests the presence of a distinct rotation curve out to 10 kpc, beyond which there is significant scatter. We show the theoretical rotation curve assuming an NFW dark matter profile with a dashed red curve. (Right) The projected \ha\ $v_{rot}/\sigma$-ratio is similar to those found in disk-like galaxies \citep{Wisnioski2015}.} 
	 \label{fig:galrot} 
\end{figure*}

\textbf{Single disk galaxy?} Assuming that \target\ is hosted by a massive rotating gas disk, we take the \ha\ kinematic maps to reconstruct the host galaxy's rotation curve, shown in Figure \ref{fig:galrot}. The observed major-to-minor axis ratio is $\sim1$, suggesting a nearly face-on orientation. Based on this elongation, the galaxy's inclination angle, is estimated to be at most $i\approx45^{\circ}$, where $i=90^{\circ}$ denotes the edge-on view. Using the relation $v_{50}=v(r)\cos\phi\sin\,i$, we calculate the rotation curve $v(r,i)$ for different inclination angles: $i=15^{\circ}$, $30^{\circ}$, and $45^{\circ}$. The median dynamical mass calculated at at $r=10\ \textrm{kpc}$ with the different $i$ is $M_{dyn}\sim10^{12}\textrm{ M}_{\odot}$. We adopt $i=30^{\circ}$ that corresponds to reasonable $v_{rot}$ values. The resultant \Mdyn\ is much higher than that of star-forming galaxies at this epoch \citep{Erb2006, Maseda2013}, which could support the merger scenario. 

We plot the de-projected rotation velocities, $v(r,i=30^{\circ})$ in Figure \ref{fig:galrot}, alongside the theoretical halo velocity curve (red curve), assuming a symmetric Navarro-Frenk-White (NFW) dark matter profile and an exponential baryonic disk component. $v(r,i)$ appears to follow the symmetric rotation curve within 10 kpc, but possibly clusters into at least two distinct rotation tracks around 5 kpcs, one at $\sim300\ \textrm{km s}^{-1}$ and another at $\sim400\ \textrm{km s}^{-1}$, albeit with significant scatter. Beyond 10 kpcs, $v(r,i)$ displays a considerable scatter, potentially influenced by merger-induced tidal debris or a warped disk. We also plot the $v_{rot}/\sigma$ ratio, which is similar to a disk-like rotation predicted by \citet{Wisnioski2015}. Thus we have a confusing interpretation of a merger-like, disk-like system. 

Interestingly, simulations of galaxy mergers across cosmic epochs predict the possibility of finding dual quasars hosted by a single galaxy ($z\sim8$ \citealt{Mayer2024}; $z\sim2$ \citealt{Dadiani2024}; and $z<1$ \citealt{RosasGuevara2019}). For instance, \citet{Mayer2024} demonstrated that a major merger at $z\sim8$ could form a nuclear supermassive disk that fragments and produces a pair of massive black holes ($10^6$ to $10^8\textrm{ M}_{\odot}$) through direct collapse within the same galaxy. Other simulations predict that a large fraction of dual quasars may reside in a single host galaxy: \citet{RosasGuevara2019} predicts this for 30\% of mergers at $z\sim0.8-1$, while \citet{Dadiani2024} finds nearly half of the dual quasars at $z\sim2$ exhibit this configuration. 

The \texttt{Astrid} simulation \citep{Dadiani2024}, which investigates the dynamical and accretion histories of dual quasars at $z\sim2$, offers relevant theoretical insights for \target. 
\cite{Dadiani2024} found that not only are dual quasars found in single galaxy systems, but the most massive pairs found in high stellar mass galaxies also produce SMBH pairs with similar properties (i.e.~mass and luminosity). Furthermore, up to 80\% of dual quasar systems are hosted by disk-dominated galaxies. Thus, the gas dynamics of \target\ observed with NIRSpec align well with these theoretical predictions, supporting the single-disk galaxy scenario, following a major merger.

\subsection{Synchronized and Enhanced accretion? }\label{sec:disc:syncBH}
While it is difficult to ascertain the exact formation history and merger stage of \target\ with the current JWST data, we further explore the observed quasar properties to understand the formation paths of the twin SMBH phenomena. 

Although empirical evidence connecting galaxy merger-driven gas inflows and quasar activity remains highly debated \citep[e.g.][]{Urrutia2008, Ellison2011, Glikman2015, Mechtley2016, Villforth2017, Ellison2019, Marian2019, Pierce2023, Breiding2024}, numerical simulations suggest a strong physical connection between major mergers and quasar triggering \citep{Hopkins2009, Torrey2020}. Following a merger involving a gas-rich galaxy, the two SMBH may be fed by the same galactic gas reservoir such that the triggered quasars may have similar accretion properties and appear like ``twins.'' Although high \eddrat\ may be allowed, the accretion may be constrained to a narrow range of \eddrat\ to look identical. The exact \eddrat\ are difficult to predict in these models since quasar accretion occurs on small scales not probed by galaxy-wide calculations. Local obscuration can also result in differences in the observed quasar properties, but \target\ shows little evidence of obscuration and dust reddening, as indicated in the rest-frame optical (\HaHb\ decrement in Figure \ref{fig:lineratio}) and X-ray \citep{ChenYC2023a}. 

A notable feature uncovered by JWST is the strong rest-frame optical \feii\ emission, likely produced at or close to the Broad Line Region \citep{Boroson1992, Hu2008}. \target\ is consistent with the Eigenvector 1 quasars \citep{Boroson1992, Boroson2002, Shen2014} that exhibit strong optical \feii, weak \oiii, and high $\lambda_{Edd}$. The prominent \feii\ may support the high $\lambda_{Edd}$ estimates in \target. The high $\lambda_{Edd}$ may suggest enhanced accretion, possibly fueled by inflows caused by the merger or dynamical interactions, although such inflows are not directly detected with NIRSpec. 

One possible formation scenario involves a merger between two galaxies with similar pre-merger properties, perhaps within the same merger tree. Another plausible scenario is a merger between a gas-rich and a gas-poor galaxy. \cite{VanWassenhove2012} even predict a correlated growth of the SMBH pair towards the later stages of the merger at $<10\textrm{ kpc}$. However, maintaining similar accretion histories over extended timescales is challenging, which may explain the rarity of ``twin'' dual quasar systems. 
 
\subsection{Comparison with low redshift analogs}\label{sec:disc:lowz}
The general evolution of dual quasars remains challenging to assess with just \target\ as a case study. However, some notable low redshift dual quasars with similar nuclear separations provide valuable context. Two well-studied examples are Mrk 463 at $z=0.005355$ \citep{Bianchi2008, Treister2018} and Mrk 739 at $z=0.02985$ \citep{Koss2011, Tubin2021}. These low redshift systems benefit from higher spatial resolution, allowing for detailed investigations of merger dynamics, including both gas and stellar components. 

Both Mrk 463 and Mrk 739 exhibit tidal tails, yet their morphologies differ significantly. Mrk 463 displays the highly distorted structure expected in a merger, while Mrk 739 appears relatively undisturbed. Detailed kinematic decomposition of Mrk 739 reveals it as a blended early-stage merger: a star-forming spiral galaxy superimposed on its companion elliptical galaxy during their first passage. Similarly, MaNGA 1-114955 at $z=0.09$, is a pair of counter-rotating disk galaxies at pre-coalescence \citep{Ciraulo2021}. MCG-03-34-64 at $z=0.016$ is a $\sim100\ \textrm{pc}$ separation dual AGN embedded in a gas-rich massive disk with little if any, morphological evidence for merger \citep{TrindadeFalcao2024}. These cases illustrate that galaxy mergers and dual quasar hosts may not necessarily require obvious morphological distortions. Detailed spatially resolved kinematic decomposition is essential to uncovering their nature. This supports the plausibility of \target\ being a merger that appears as a single disk galaxy, as explored in Section \ref{sec:disc:merger}.  

Another intriguing aspect of \target\ is the detection of two SMBHs with similar spectral properties explored in Section \ref{sec:disc:syncBH}. Empirical evidence of SMBH pairs with similar spectral properties exists in low-redshift systems. For example, \cite{Koss2023} discussed UGC 4211, a $z=0.03474$ dual AGN at 230 pc separation, with optical spectra indicative of similar SMBH masses. UGC 4211 is a confirmed dual quasar in a merger with measured velocity offsets between the two quasars reaching $\sim150$ \kms. However, unlike UGC 4211, \target\ lacks clear merger signatures, adding to the ambiguity of its nature. 

\target\ lacks evidence for significant quasar-driven outflows. The absence of nuclear winds and extended, broadened \oiii\ may indicate that quasar feedback via outflows is not (yet) a significant factor.  \cite{Ruby2024} detected a dual quasar, Mrk 266, at $z=0.028$, which hosts a small and young outflow. This observation may imply that we would not likely see powerful extended outflows in dual quasars at high redshift. Mrk 266 provides evidence that mergers may trigger star formation and quasar activity. 

Similarly, the minimal obscuration in \target\ could imply that the quasar separation is too large for extreme (near/super-Eddington) accretion, often associated with obscured phases, to dominate. This contrasts with theoretical predictions \citep{Blecha2018} and observational studies of local dual quasars such as Arp220 \citep{Veilleux2009a}, which report increased obscuration at smaller separations. The difference may stem from the VODKA's preference for selecting non-dusty, unobscured quasars, potentially leaving a population of obscured dual quasars waiting to be discovered \citep[e.g.][]{Koss2018,Barrows2023}. 

Although no direct low redshift analogs for \target\ are currently known, these examples demonstrate that dual quasar and their associated galaxy mergers display diverse properties. This complicates follow-up studies of dual quasar candidates, particularly at higher redshifts where spatial resolution and data quality are affected. Unfortunately, it is difficult to perform similar kinematic decompositions of \target\ with our current JWST data. Future studies may resolve this ambiguity by comparing the kinematics and morphologies of the stellar component with the ionized emission line components.

\section{Constraints of VODKA}\label{sec:disc:vodka}
Lastly, we comment on the VODKA technique for selecting dual quasar candidates. Selection effects may contribute to the kinds of quasar pairs discovered, such as \target\ with two similar black hole properties. \cite{Hwang2020} selected \gaia\ sources brighter than $G<19\textrm{ Vega mag}$ to measure intrinsic variability to within a few percent. Since the magnitude selection is determined by the brightness limit of \gaia, VODKA is more efficient at uncovering intrinsically luminous $z>2$ pairs due to decreased contamination by the host galaxy (e.g.~offset AGNs) and decreased lensing probability \citep{Hwang2020, Shen2021}. Moreover, VODKA may preferentially select quasar pair candidates with similar fluxes. If one quasar's flux dominates or the two black hole masses are too different, then the light centroid would be stable and centered on the brighter source, and VODKA may not detect any strong astrometric noise. These VODKA-selected pairs may be lensed quasars \citep{Li2023, Gross2023} or physical dual quasars with equal fluxes. Distinguishing lensed and physical pairs requires extensive multiwavelength follow-up studies \citep{Gross2023}. 

It appears that the intrinsic abundance of quasar pairs, both lensed and physical pairs, at cosmic noon is low \citep{Shen2023a}. Currently, there are disagreements on the abundance of dual quasars across cosmic time \citep[e.g.][]{Silverman2020, Tang2021, Shen2023a, Perna2023}, when comparing observations with predictions from hydrodynamical simulations \citep[e.g.][]{DeRosa2019, Volonteri2022}. Disentangling the division between lensed quasars and physical pairs remains a challenge \citep{Shen2023a}, especially in the sub-arcsec regime. Furthermore, the properties of the dual quasar population remain poorly understood. It is currently difficult to determine if the discovery of close-separation dual quasars with similar quasar properties like \target\ is statistically significant or is simply a selection effect. 

On the other hand, studies by \cite{Silverman2020, Tang2021} suggest that dual quasars have diverse properties. There are even some intriguing cases of a dual 6-kpc separation dual quasar that is quadruply lensed \citep{Lemon2022}. Although these cases may only encompass a minority of the dual and lensed populations, these peculiar discoveries accentuate the complexity and importance of detailed follow-up studies. This highlights the necessity for deep, spatially resolved, multi-wavelength follow-up observations to confirm the nature of dual quasars.

\section{Summary and Conclusion} \label{sec:concl}
In this paper, we present JWST NIRSpec IFU observations of \target, a dual quasar candidate at $z=2.17$ with a physical separation of $3.8\ \textrm{kpc}$. \target\ was identified using the VODKA method. With JWST, we report the first direct detection of the faint host galaxy surrounding the two quasars. We analyze the NIRSpec IFU observations with \qdfit, which fits the datacube with the quasar, polynomial continuum, and emission line models to decompose the spectra into the quasar PSFs and host galaxy. Previous multiwavelength imaging and spectroscopic observations of \target\ (i.e.~Gemini, HST, Chandra, VLBA) spatially resolved the two quasars (\targNE\ and \targSW) and noted an indirect detection of tidal tails, potentially from their host galaxies \citep{ChenYC2023a}. With JWST, we confirm and detect the extended ionized gas, best traced by the narrow \ha\ emission lines.

We extracted the aperture spectra of each of the two quasars, \targSW\ and \targNE, and fit them with \pyqsofit. Surprisingly, both quasars exhibit similar spectral properties, including comparable emission line withs, continuum shapes, and prominent \feii\ line complexes. Their bolometric luminosities exceed $10^{46}\textrm{ erg s}^{-1}$, and single-epoch black hole masses estimates are $\sim10^9\ M_{\odot}$, with relatively large \eddrat. Despite the observed similarity, subtle spectral differences are apparent, most notably a velocity offset between the quasars of $\Delta v \sim200\textrm{ km s}^{-1}$. This velocity offset, combined with the detection of the host galaxy at the same redshift as the quasars, disfavors the lensing hypothesis. However, the strong lensing scenario cannot be excluded, especially due to uncertainties in $\Delta v$ and the observed quasar similarities. 

Using \qdfit\ we decompose the two quasar spectra from the faint gas emission corresponding to the host galaxy and simultaneously fit the emission lines and continuum. We derive intensity, velocity offset, and velocity dispersion maps for \ha, \hb, \oiii, \nii, \sii, and \oi. The narrow line \ha\ emission extends nearly $2.5\arcsec$ or $\sim20\ \textrm{kpc}$ in diameter, while \hb\ and \oiii\ emission are more compact, primarily centered around \targNE. No quasar-driven outflows are detected with the narrow \oiii\ emission. Optical line ratio diagnostics based on \niiha, \siiha, \oiha, and \oiiihb\ indicate a mix of ionization sources, with quasar photoionization near the two quasars and significant star formation. The total \ha\ flux associated with star formation is $6.0\times10^{-15}\ \textrm{erg s}^{-1}\ \textrm{cm}^{-2}$, which translates to $\textrm{SFR}\sim1,700\ M_{\odot}\textrm{yr}^{-1}$, consistent with MIR results \citep{ChenYC2024}.

Surprisingly, the ionized \ha\ gas kinematics reveal a large, disk-like rotation curve, rather than an asymmetric profile expected from an ongoing merger. The $v_{50}$ distribution is mostly symmetric. Assuming the \ha\ gas traces a single galaxy, we estimate $M_{dyn}\sim10^{12}\ M_{\odot}$. We explore possible scenarios to explain the nature of \target. Assuming \target\ is not a lensed quasar, \target\ could be either an ongoing gas-rich merger, a single disk galaxy hosting two luminous quasars, or a combination of both. The similarity of the quasars may also suggest a ``synchronized'' accretion scenario in which the two black holes are fueled simultaneously, possibly by the host galaxy dynamics. Intriguingly, these scenarios are supported by simulations such as \cite{Daddi2021}. Observations of low redshift analogs, including dual quasars and galaxy mergers, ``disguised'' as a single galaxy further underscore the difficulty of disentangling these possibilities. 

This JWST study highlights the power and utility of integral field spectroscopy using JWST, especially in detecting the diffuse, extended gas associated with the host galaxies at cosmic noon. As one of the first studies to probe the gas dynamics of dual quasars at this epoch, JWST reveals new insights previously inaccessible. Although VODKA effectively identifies dual quasar candidates, we find that the VODKA methodology may preferentially select quasar pair candidates with similar fluxes, leading to the discovery of sources like \target\ with similar black hole properties (\Mbh\ and \Lbol). If \target\ is confirmed as a dual quasar, it remains uncertain whether its properties are representative of the dual quasar evolution or a rare subset of the dual quasar population. This study also highlights the difficulty of ruling out the gravitationally lensing hypothesis for small-separation dual quasar candidates, emphasizing the need for exhaustive multi-wavelength, spatially resolved observations of the quasars and their host galaxies.

\begin{acknowledgments}
This work is based on observations made with the NASA/ESA/CSA James Webb Space Telescope. The data were obtained from the Mikulski Archive for Space Telescopes at the Space Telescope Science Institute, which is operated by the Association of Universities for Research in Astronomy, Inc., under NASA contract NAS 5-03127 for JWST. These observations are associated with program \#\progNum. Some/all of the data presented in this article were obtained from the Mikulski Archive for Space Telescopes (MAST) at the Space Telescope Science Institute. The specific observations analyzed can be accessed via \dataset[DOI]{https://doi.org/10.17909/jnee-4g18}.

Support for program \#\progNum\ was provided by NASA through a grant from the Space Telescope Science Institute, which is operated by the Association of Universities for Research in Astronomy, Inc., under NASA contract NAS 5-03127. This work is supported by the Heising-Simons Foundation and Research Corporation for Science Advancement, and NSF grant AST-2108162 (X.L., Y.S., Y.-C.C., A.G.). D.S.N.R. was supported under program \#DD-ERS-1335 from the Space Telescope Science Institute under NASA contract NAS 5-03127.

The authors thank the anonymous referee for their helpful suggestions that vastly improved this work. Y.I. thanks D.~Coe, A.-C.~Eilers, E.~Glikman, T.~Heckman, S.~Knabel, C.~Norman, D.~Neufeld, M.~Onoue, J.~Silverman, M.~Stiavelli, Y.-C.~Taak, and M.~Yue for useful discussions.

\end{acknowledgments}

%

\facilities{JWST (NIRSpec)}


\software{\texttt{astropy} \citep{astropy2013,astropy2018,astropy2022}, \qdfit\ \citep{q3dfit2023}, \pyqsofit\ \citep{GuoH2018} }

\bibliography{vodka_v2}{}
\bibliographystyle{aasjournal}

\end{document}